\documentclass[reprint,showpacs]{revtex4-1}
\usepackage{graphicx}
\usepackage{multirow}
\usepackage{epsfig}
\usepackage{color}

\newcommand{\bea}{\begin{eqnarray}}
\newcommand{\eea}{\end{eqnarray}}
\newcommand{\ba}{\begin{array}}
\newcommand{\ea}{\end{array}}

\newcommand{\be}{\begin{equation}}

\newcommand{\ee}{\end{equation}}
\newcommand{\bt}{\begin{teo}}
\newcommand{\et}{\end{teo}}

\newcommand{\la}{\lambda}
\newcommand{\om}{\omega}

\begin{document}

\title[Parametrically kicked Hamilton systems]
{Statistical properties of 1D parametrically kicked Hamilton systems}

\author{Dimitris Andresas, Benjamin Batisti\'c and Marko Robnik}

\address{CAMTP - Center for Applied Mathematics and Theoretical Physics,
University of Maribor, Krekova 2, SI-2000 Maribor, Slovenia, European Union}

\email{dimitraklos@hotmail.com, Benjamin.Batistic@gmail.com, Robnik@uni-mb.si}

\begin{abstract}
We study the 1D Hamiltonian systems and their
statistical behaviour, assuming the initial microcanonical distribution
and describing its change under a parametric kick, which by definition
means a discontinuous  jump of a control parameter of the system.
Following a previous work by Papamikos and Robnik {\em J. Phys. A:
Math. Theor. {\bf 44} (2011) 315102} we specifically analyze the
change of the adiabatic invariant (the action) of the system under
a parametric kick: A conjecture has been put forward that the
change of the action at the mean energy always increases, which
means, for the given statistical ensemble, that the Gibbs entropy
in the mean increases. By means
of a detailed analysis of a great number of case studies we show
that the conjecture largely is satisfied, except if either the 
potential is not smooth enough, or if the energy is  too close to a
stationary point of the potential (separatrix in the phase space). 
Very fast changes in a time dependent
system quite generally can be well described by such a picture and by
the approximation  of a parametric kick, if the change of the parameter 
is sufficiently fast and takes place on the time scale 
of less than one oscillation period. We discuss our work in the
context of the statistical mechanics in the sense of Gibbs.
\end{abstract}

\pacs{05., 05.45.-a, 05.45.-Ac}

\maketitle

\section{Introduction} \label{Intro}

In a recent work Papamikos and Robnik \cite{PapRob2012} have studied
time dependent nonlinear Hamiltonian oscillators from the point of view
of their statistical properties, in order to generalize a series of
studies on the time dependent linear oscillator by Robnik
and Romanovski \cite{RR2006a,RR2006b,RRS2006,KR2007,RR2008},
where the rigorous WKB method has been employed \cite{RR2000}. We 
are interested in the time evolution of a microcanonical
ensemble of initial conditions. If the evolution is ideal adiabatic
(i.e. infinitely slow), then the adiabatic invariant, which is also
the action of the system, or the area inside the contour of
constant energy in the phase space (divided by $2\pi$), is conserved, 
and this is precisely the adiabatic
theorem on one-dimensional Hamiltonian systems \cite{Arnold}, provided
we do not cross a separatrix during the adiabatic process.
What happens if the changing of the system parameter is not adiabatic?
For the linear oscillator with an arbitrary time dependence it has been
shown rigorously in the above mentioned papers (see in particular the review
\cite{RR2008}) that the value of the adiabatic invariant at the
average value of the energy during the evolution is always
increasing. Since the adiabatic invariant is proportional to the
number of states, this implies an irreversibility in the mean,
because the entropy is the logarithm of the number of states,
in the sense of statistical mechanics, as explained in section 2.
This finding was a motivation to analyze the {\em nonlinear}
oscillators from this point of view. In \cite{PapRob2012} it was
shown using the numerical techniques (highly accurate symplectic
integrators of 8th order, \cite{M1995,McL2002,HLW2006,LeimReich,SenzSernaCalvo,
SY1996,Y1990,Y1993,P2011}) 
that for slow but {\em not} adiabatic
changes in a quartic oscillator the adiabatic invariant at the
mean energy can {\em decrease}, just due to the nonlinearity and
nonisochronicity. However, for sufficiently fast changes the property
is restored, especially in the extreme case of a {\em parametric kick},
when the system's parameter jumps discontinuously. This led us
to demonstrate analytically by a rigorous calculation for the case
of homogeneous power law potentials (of which the quartic
oscillator and the harmonic oscillators are two special cases), 
that the adiabatic invariant at the average final energy 
indeed increases under a parametric kick. 

Therefore, we \cite{PapRob2012} have put forward a Conjecture 
(henceforth called {\em PR Conjecture}) that the adiabatic invariant for 
an initial  microcanonical ensemble at the mean
energy always increases under a parametric kick. The purpose of
the present work is to investigate the validity and the conditions
under which it is true. We show by a series of case studies, that
the PR Conjecture is largely true, except  if either the 
potential is not smooth enough, or if the energy is  too close to a
stationary point of the potential (close to a separatrix in the phase
space). The latter complications are
not unexpected, because existence of a separatrix in the phase space
always complicates matters, e.g. implies violation of the adiabatic 
theorem, because the averaging method does not work there, since
the period of oscillation is infinite. Also, breaking the smoothness 
properties of the potential obviously can break
"communication" between the different parts of the potential well. 
A very recent review of the main ideas has been published in \cite{Rob2013}.

In a more general context, time dependent Hamiltonian systems 
are very interesting and important dynamical models, 
where many interesting questions  about their
statistical behaviour can be studied \cite{Arnold, Lochak, Zaslavsky, Ott}.
The time dependence of the Hamilton function describes, or models, the
interaction of the system with the environment.
Whilst the energy of the system is not conserved, 
the Liouville theorem of course
still applies and thus the phase space volume is preserved by the flow. 
One of the major questions is the time evolution of the energy of certain 
ensembles of initial conditions. The microcanonical ensemble is the
most fundamental, like in statistical mechanics, and we 
investigate how it develops in time. In the ideal adiabatic processes,
which are infinitely slow, the adiabatic invariant is conserved,
and using this conservation law we can calculate the sharply
defined energy changing in time. 
If the process is non-adiabatic, having a finite
speed of changing the Hamilton function, the energy will be spread
around its mean value.
In the linear oscillator \cite{RR2006a}-\cite{RR2008} the distribution function
of the energy is universal, independent of the driving law of the frequency as
a function of time, and  is given by the arcsine distribution.
In nonlinear systems this universality is lost, and the evolution
of the energy distribution can exhibit a rich variety of behaviour. 
Once we have the energy distribution, we can calculate distribution
of other dynamical quantities, in particular of the adiabatic
invariant. 

In particular in time periodic (Floquet) systems 
we can find a very rich behaviour, from integrability to full chaoticity
(ergodicity), and also the scenario in between, namely the case of a mixed phase
space, even in 1D systems. One example is the kicked rotator 
(standard map) and many other time periodic systems \cite{Chirikov,Zaslavsky}. 
More recent works included the periodic parametric kicking of
the quartic oscillator \cite{PapRob2012} and the periodic linear
driving (sawtooth driving law) of the quartic oscillator
\cite{PapSowRob2012}.

Time periodic systems
are interesting also from the point of view of the Fermi acceleration,
including the quantum mechanical counterparts, 
that is unlimited growth of the energy, in 1D systems and higher dimensions.
For some recent works see \cite{BatRob, BatRob2012, Sch2009, Schmelcher} 
and references therein. Lots of interesting empirical
material has accumulated, including the power law behaviours with
universal scaling properties \cite{LMS2004}. 

In this paper we study the parametrically kicked 1D Hamiltonian systems, 
trying to work out conditions under which the PR Conjecture holds true,
showing, as mentioned above, that it is largely satisfied except when
the initial energy (of the microcanonical ensemble) is too close to
a stationary point of the potential (separatrix in the phase space),
or if the potential is not analytic and not sufficiently smooth.
For these reasons we shall speak of the PR property, namely that
a certain potential behaves in agreement with the PR Conjecture,
and thus possesses the PR property, but possibly only for a certain 
range of energies, or entirely not. 

The paper is organized as follows.
In section 2 we explain the connection  to the statistical
mechanics in the sense of Gibbs, 
especially for small number of degrees of freedom, explaining
why the Gibbs entropy is fundamental and correct 
(in contradistinction to the Boltzmann entropy), 
as emphasized very recently also by 
Dunkel and Hilbert \cite{DunHil2014},
corroborating the views of Gibbs \cite{Gibbs1902}, Einstein \cite{Ein1911}
and Hertz \cite{Her1910}.
In section 3 we present the general theory of parametric kicks, 
in section 4 we analyze 
the examples where the PR Conjecture is entirely satisfied, in section 5 we
analyze the counter examples, where the PR Conjecture is violated
or partially violated (that is, its validity applies only to a certain
energy range of the potential). In section 6 we discuss the results 
and conclude. In Appendix A as an overview we summarize the list of potentials
with valid or broken PR property.

\section{The PR-property and its connection to the statistics
in the sense of Gibbs}

In statistical mechanics of classical mechanical systems the
most fundamental ensemble to calculate the entropy, and thus
all other equilibrium properties, is the Gibbs microcanonical
ensemble, based on the number of states $\Omega(E)$ {\em inside 
the (closed) energy  surface} of energy $E$, calculated as the
phase space volume

\be  \label{Omega}
\Omega (E) =  \int_{H(q,p)\le E} d^fq\;d^fp.
\ee
Here $f$ is
the number of degrees of freedom. Following Gibbs the entropy,
called {\em Gibbs entropy}, in order to distinguish it from other
definitions like, e.g. the Boltzmann entropy,
is defined as follows

\be \label{Gibbsentr}
S_G (E) = k_B\ln \Omega (E),
\ee
where $k_B$ is the Boltzmann constant. $\Omega$ has the dimension of
the $2f$-dim phase space volume, which is a technical nuisance when taking
the logarithm of it. This difficulty can be removed by dividing
$\Omega$ by a constant $a$ with the same physical dimension.
However, so long as we are interested only in differences of the
entropy, which is the case in the classical statistical
mechanics, such a constant $a$ drops out from all calculations and
thus has no physical significance.
Usually, however, the natural choice is $a=(2\pi\hbar)^f$, thus making
the definition of $S_G$ compatible with the quantum version,
where the entropy is well defined in absolute terms. In this 
paper we deal only with classical mechanics.

The fundamental role of $\Omega$     
has been established by Gibbs himself \cite{Gibbs1902},
later discussed by Hertz \cite{Her1910} and Einstein \cite{Ein1911},
and recently corroborated in a critical analysis by Dunkel
and Hilbert \cite{DunHil2014}, showing that Gibbs entropy is at variance
with the Boltzmann entropy. The latter is defined as the logarithm
of the number of states inside an {\em energy shell} around the energy $E$, 
and differs from the Gibbs entropy, especially
in systems with a small number of degrees of freedom $f$ (small systems),
such as treated in this paper.
It is precisely the Gibbs entropy which gives the
right answers and results in small systems. For example, in the case of
an ideal monoatomic gas it was shown in \cite{DunHil2014}, regarding
e.g.  the calculation of the equipartition, that Gibbs definition
is the right one, whilst the Boltzmann
entropy differs at small $f$, but of course nevertheless agrees with the Gibbs 
entropy for large systems (large $f$). 

It has been realized by Gibbs \cite{Gibbs1902}, Hertz \cite{Her1910} 
and Einstein \cite{Ein1911}, that the fundamental quantity of
classical statistical mechanics is $\Omega(E)$, defined in
(\ref{Omega}).  It is precisely the adiabatic invariant of the
system, which is conserved under adiabatic infinitely slow changes,
as proven by Paul Hertz \cite{Her1910} for ergodic systems. 
Also quantum mechanically,
$N(E) = \Omega/a$ is precisely the number of quantum eigenstates of a
bound system below the energy $E$, which is also the adiabatic
invariant in quantum mechanics.
In one degree of freedom systems, $f=1$, we have of course
$\Omega (E) = 2\pi I (E)$, where $I(E)$ is the classical Hamilton 
action of the system at energy $E$. Hence the importance of $I(E)$,
which we also call adiabatic invariant.

On the other hand, if the system (its Hamilton function) depends
on time nonadiabatically, meaning having finite speed of changing
the parameter, the energy of the system still changes, but now it
has a distribution around its mean value $\bar{E}$, and $\Omega(E)$ 
also depends on time,  having a certain distribution,  
but the most interesting and important question is then under
what conditions $\Omega (\bar{E})$ will increase or decrease in time,
implying increasing or decreasing Gibbs entropy (at the mean energy), 
respectively.

What is the relevance of 1D Hamiltonian systems in this context?
If we have an ensemble, even macroscopic ensemble, of identical
1D {\em noninteracting} systems, the behaviour of the macroscopic
ensemble will be obviously very much determined by the behaviour of a single
system. One example is the ideal gas. Enclosing particles 
in a 1D box of length $L$, 
the behaviour of the 1D gas is governed by the behaviour 
of one particle interacting with the moving walls, simply due
to the absence of interactions between the particles. By calculating
$\Omega = 2Lp = 2L \sqrt{2mE}$ for a single particle, we find 
immediately  $E= \frac{1}{2}k_B T$, if $1/T$ is interpreted as
$dS_G/dE$. This is precisely the equipartition law. Since
$E$ and $S_G$ are additive, it follows immediately $E=nk_BT/2$
for $n$ noninteracting particles in such  1D box. The
"temperature" $T$ here can be understood also as the time average
of the particle's kinetic energy over sufficiently many
oscillation periods, divided by $k_B/2$. All equilibrium properties
of the 1D ideal gas can be determined in that way, using $S_G$. 
Moreover, if $L(t)$ is a function of time, we can calculate
the evolution of the energy $E$, for an ensemble of initial
conditions, and $\Omega$.  
This picture can be generalized to the 3D ideal gas
whose thermodynamic equation of state can be derived.
Similar approach is applicable to the general
time dependent 1D nonlinear Hamiltonian oscillators.

Of course, the most fundamental is the {\em microcanonical
ensemble} of initial conditions, all located on the initial
energy contour with sharp energy $E_0$, but distributed uniformly
on the torus w.r.t. the canonical angle (the phase). Then, $E$ 
changing in time is a function of the initial condition, 
and spreads around its mean value $\bar{E}$, also varying
in time. Calculating the $\Omega (\bar{E}) = 2\pi I(\bar{E})$
at the mean energy then enables us to calculate the Gibbs 
entropy $S_G = S_G(\bar{E})$ as a function of time, and with it thus all
the "thermodynamic" properties. By the additivity of $S_G$ 
such a picture can be generalized
to arbitrary 1D time dependent and {\em noninteracting} 
Hamiltonian systems.

It has been shown by Robnik and Romanovski 
\cite{RR2006a,RR2006b,RRS2006,RR2008} that in the case of
the 1D linear oscillator with {\em arbitrary} parametric
driving (the frequency $\omega(t)$ is an arbitrary function
of time, including a discontinuous jump (=parametric kick))
the $\Omega (\bar{E})$ always increases, and so does the
Gibbs entropy $S_G(\bar{E})$, except in adiabatic, infinitely slow
processes, where it is exactly conserved.  In nonlinear 1D
systems this property is in general lost, just due to the nonlinearity.
Thus $\Omega (\bar{E})$ and $S_G(\bar{E})$ can decrease.
For the adiabatic processes $\Omega(\bar{E})$ is of course exactly conserved, 
by the theorem on adiabatic invariants \cite{Arnold}, as proven
by Paul Hertz \cite{Her1910} for the general ergodic systems of any
degrees of freedom, thus including $f=1$. But for slow
although not adiabatic processes $\Omega(\bar{E})$ can decrease, just due
to the nonlinearity, which was demonstrated by Papamikos
and Robnik, as mentioned above. This is an important observation,
because it indicates that the nonlinear interactions can
lead to the decrease of the entropy of parametrically driven systems, 
which is not possible in the linear oscillators. Nevertheless,
for sufficiently fast parametric driving, we intuitively expect
that the PR property holds true and thus 
the law of the increasing entropy is restored.
This is precisely what we observe, and the extreme case is the
case of parametric kicking, being the fastest possible change. 
This is the main motivation
of the present work, where we systematically explore by
a number of case studies when the PR property is valid
or not. One should bear in mind that the 
parametric kicking is a good, leading, approximation
of systems which undergo very fast changes of the parameter,
on time scales of less than one oscillation period,
as has been also demonstrated by Papamikos and Robnik
mentioned above.

Of course, there is a number of open questions, 
in fact a whole programme of research in this direction:
What can we say about general parametric driving of nonlinear
1D Hamiltonian systems? How can we treat {\em collectively} 
an ensemble of noninteracting identical 1D nonlinear driven 
oscillators, by calculating $\Omega$ as a function of time?
Furthermore, how can we generalize all results for 
driven higher dimensional oscillators, first for
a single system, and then collectively for an ensemble 
of identical noninteracting systems?

It turns out that already the first step in this programme,
namely the case of the parametrically kicked 1D systems, 
is difficult enough, dealt with in this paper.

\section{General theory of parametric kicks in 1D Hamiltonian systems} \label{General}

We consider Hamiltonian systems with one degree of freedom in the form
quadratic kinetic energy (except for the subsection \ref{hompot}, where
the kinetic energy is more general) plus potential, where the potential 
depends linearly on the parameter $A$, which is the system's control 
parameter,

\be \label{generalhamiltonian}
H(p,q) = \frac{p^{2}}{2} - Af(q),\quad A>0. 
\ee
Sometimes we shall use also the notation for the potential
$V(q)=-Af(q)$. The time dependence that we study is an instantaneous jump of
$A$, from $A_0$ to $A_1$. Thus, the initial and the final
Hamilton functions are

\bea \label{inifinHam}
 H_0= H(q,p,A_0) &=& \frac{p^{2}}{2} - A_0 f(q), \nonumber\\
H_1= H(q,p,A_1) &=& \frac{p^{2}}{2} - A_1 f(q).
\eea
We assume the {\em microcanonical ensemble of initial conditions}, which 
means that the initial energy $E_0$ is sharply defined, and the distribution
of the initial conditions is uniform with respect to the canonical
angle variable $\theta$, which means that the density 
of points on an infinitesimal
interval on the energy contour $H_0=E_0$ is proportional to the length
of time spent in that interval under the dynamics of the initial but
frozen Hamiltonian $H_0$. Thus, for an observable $F(q,p)$, the 
final $(q,p)$ are functions of $E_0$, $\theta$ and time, and the 
average at time $t$ is defined as follows

\be \label{average}
\langle F \rangle (E_0,t) =  \frac{1}{2\pi} 
\int_0^{2\pi} F\left(q(\theta,t),p(\theta,t)\right) d\theta.
\ee
This can be also written as (suppressing the arguments) 

\be \label{timeaverage}
\langle F \rangle = \frac{1}{T_0} \oint F(\tau)\, d\tau = 
\frac{\oint F(q,p)\, dq_0/p_0}{\oint dq_0/p_0},
\ee
where $T_0=\oint dq_0/p_0$ is the period of the oscillation of the
initial frozen system, $\tau$ is its time, and $q,p$ are
regarded as functions of the initial point $q_0,p_0$ on
the energy contour $E_0$, and of time $t$.

When the jump $A_0\rightarrow A_1$ takes place,
the coordinate in the configuration space  $q$ and the canonical
conjugate momentum $p$ remain continuous, because $q$ is by definition
a continuous variable, whilst $p$ is continous because there is no external
kick (Dirac delta function peaked force) 
acting on the system. These statements are not
trivial, because we cannot choose for $q,p$ the action-angle variables,
simply because they are not defined in time dependent systems. Indeed,
if $p,q$ were action-angle variables $(I,\theta)$, not changing at all,
nothing at all would happen in the system, because $H(I,\theta)$ 
would not change at all. 
Thus, it is important to realize that we must work in the ordinary
phase space $(q,p)$. Also, we must remark that the phase flow
in such case is only $C^0$.

For the final energy $E_1$, after the jump, we can write

\bea \label{finE1}
E_1 &=& \frac{p^{2}}{2} - \frac{A_1}{A_0} A_0 f(q) \nonumber\\
&=& \frac{p^{2}}{2} + \frac{A_1}{A_0} (E_0 - \frac{p^{2}}{2}) \nonumber\\
&=& \frac{p^{2}}{2}(1-x) + E_0 x ,
\eea
where we shall use the notation $x=A_1/A_0$ throughout this paper.
So the final energy $E_1$ has some distribution implied by the nature
of the initial microcanonical ensemble and the change of the geometry
of the phase space, and is now only a function of initial and final
$p$ at fixed $x$. We can thus immediately calculate the average value of
$E_1$, denoted by $\langle E_1 \rangle$, namely

\bea \label{aveE1-0}
\langle E_1 \rangle &=& (1-x)\frac{\langle p^{2} \rangle}{2} + E_0 x 
= E_0 x + \frac{(1-x)}{2T_0} \oint p^{2} \; dt \nonumber \\
&=& E_0 x + \frac{(x-1)}{2T_0} \oint p \; dq 
= E_0 x + \frac{(1-x)\pi}{T_0}I_0,
\eea
where the integration $\oint$ is taken over the entire oscillation cycle,
that means from the smaller turning point usually denoted by $q_1$
to the larger one $q_2$, and back.
$T_0 = T(E_0,A_0)$ is the period at the initial energy $E_0$ and 
$I_0 = I(E_0,A_0) $ is the action as a function
of the energy $E_0$ and parameter $A_0$. They are generally defined as

\be \label{Idef}
I(E,A) = \frac{1}{2\pi} \oint p\,dq = \frac{1}{\pi}\int_{q_1}^{q_2} dq\, \sqrt{2(E+A f(q))},
\ee
and the period of oscillation $T(E,A)$ as

\bea \label{Tdef}
T(E,A) &=& \oint dt = \oint \frac{dq}{p}
= 2\int_{q_1}^{q_2} \frac{dq}{\sqrt{2(E+A f(q))}} \nonumber \\&=&
2\pi \frac{\partial I(E,A)}{\partial E}.
\eea
In the following we denote also $I_1=I(E_1,A_1)$ and $T_1=T(E_1,A_1)$.
We can also calculate $\langle E_1 \rangle$ in terms of the new action $I_1$
and period $T_1$ evaluated at $E=\langle E_1 \rangle$ and $A_1$, namely
as follows

\be  \label{aveE1-1}
\langle E_1 \rangle =  \frac{\langle p^2 \rangle}{2}  +  \langle 
-A_1f(q) \rangle_1 =
\pi \frac{I_1}{T_1} -A_1 \langle f(q) \rangle_1
\ee
where the averaging is now over the contour $E_1=H(q,p,A_1)$, and thus

\be  \label{avefq}
\langle f(q)\rangle_1 = \frac{1}{T_1} \oint_{\langle E_1 \rangle} f(q)\,dt.
\ee
Using the two expressions for $\langle E_1 \rangle$ 
we arrive at the expression for the final action $I_1$ at the average
final energy

\be \label{finI}
I_1 = I\left(\langle E_1 \rangle, A_1\right)= \frac{T_1}{T_0}I_0 (1-x) + \frac{T_1 E_0 x}{\pi} 
+ \frac{T_1 A_1\langle f(q) \rangle_1}{\pi},
\ee
and the average $\langle f(q) \rangle_1$ in (\ref{avefq}) can be further expressed 

\be \label{avefq1}
\langle f(q) \rangle_1 = 
\frac{1}{T_1} \oint_{\langle E_1 \rangle} \frac{f(q) \; dq}{\sqrt{2\left(E + Af(q)\right)}} =
\langle f(q) \rangle_1 =  \frac{2\pi}{T_1}\frac{\partial I_1}{\partial A_1}.
\ee
Therefore the final action at the final average energy from (\ref{finI}) 
is 

\be \label{finIE}
I_1 = \frac{T_1}{T_0}I_0 (1-x) + \frac{T_1 E_0 x}{\pi} + 
2 A_1 \frac{\partial I_1}{\partial A_1}\vert_{\langle E_1 \rangle}.
\ee
We can also express the action $I_1$ as, 
\be \label{finI1C}
I_1 = \frac{T_1 \langle E_1 \rangle}{\pi} + 2 A_1 \frac{\partial I_1}{\partial A_1}\vert_{\langle E_1 \rangle}.
\ee
{\bf The Papamikos-Robnik Conjecture} ({\bf PR Conjecture}) is now formulated
as $I_1 \ge I_0$, where $I_1$ is given either in (\ref{finI1C}),
or in (\ref{finIE}), or in (\ref{finI}), or more explicitly

\bea \label{PRC}
&&I_1 = I\left( \langle E_1 \rangle, A_1 \right) = I\left( \frac{\pi}{T_0}I_0 (1-x) + E_0 x,\; xA_0 \right) \nonumber \\
&&I\left( \langle E_1 \rangle, A_1 \right)\geq I\left( E_0,A_0\right),
\eea
which must be satisfied for 
all values of $E_0$, $A_0\ge 0$ and $x=A_1/A_0\in[0,\infty)$.
\\\\
If the PR property (\ref{PRC}) is satisfied only for some energy
range $E_0$ of the potential, we shall say that the PR Conjecture is 
partially sastisfied, and if is entirely violated, we shall say
that the potential does not possess the PR property.

Since $x$ is any positive real number, and since in the case
 $x=1$ nothing happens
at all (no parametric kick at all), (\ref{PRC}) must be equality
$I_1 = I_0$ for $x=1$, and strict inequality for all other $x$. This Conjecture
is difficult or almost impossible to prove in general with
available techniques, especially as it is valid
under restricted conditions. Nevertheless, we shall prove it rigorously
by direct calculations in a very large class of specific systems (potentials),
treated in section \ref{PRCY}.

Nevertheless, we can do much more in the local analysis, in the sense of 
investigating when (\ref{PRC}) has a minimum at $x=1$. We define
$x= 1 + \epsilon$, where $0< \vert \epsilon \vert \ll 1$, and look
at the Taylor expansion in $\epsilon$,

\be \label{Taylor}
I(E,A) = I_0 + L\; \epsilon + Q \; \epsilon^2 + O(\epsilon^3),
\ee
where the linear term $L$ is, using $\frac{\partial I}{\partial E} =T_0/(2\pi)$,

\be \label{L}
L = \frac{E_0T_0}{2\pi} -\frac{I_0}{2} + 
 A_0 \frac{\partial I}{\partial A}\vert_{(E_0,A_0)},
\ee
and the quadratic term $Q$ is equal to:
 
\bea \label{Q}
Q&=&\frac{1}{2}\left\{\frac{\partial^{2}I}{\partial E^{2}}\left(E_{0} - \frac{\pi I_{0}}{T_{0}}\right)^{2}
\right.
\nonumber\\
&+& \left. 2A_{0} \frac{\partial^{2}I}{\partial E \partial A}\left(E_{0} - 
\frac{\pi I_{0}}{T_{0}}\right) + A_{0}^{2} \frac{\partial^{2}I}{\partial
  A^{2}}\right\},
\eea
where all partial derivatives must be taken at $E=E_0$ and $A=A_0$.
Now we evaluate $L$ and $Q$. We prove that $L=0$, meaning that
$x=1$ is a stationary point. The calculation is
straightforward as follows

\bea \label{L=0}
A \frac{\partial I}{\partial A} &=& \frac{1}{2\pi}\oint \frac{A f(q)}{\sqrt{2E + 2Af(q)}}\; dq = \frac{1}{2\pi}\oint \frac{-E + \frac{p^{2}}{2}}{\sqrt{2E + 2Af(q)}}\; dq \nonumber\\
 &=& -\frac{ET}{2\pi} + \frac{1}{4\pi}\oint p\; dq = -\frac{ET}{2\pi} + \frac{I}{2}.
\eea
Now we calculate $Q$ in (\ref{Q}). First we calculate the second partial derivatives

\bea  \label{2ndder}
\frac{\partial^{2}I}{\partial E^{2}} &=& \frac{1}{2\pi} \frac{\partial T}{\partial E} ,\\
\frac{\partial^{2}I}{\partial E \partial A} &=& -\frac{T}{4\pi A} -\frac{E}{2\pi A}\frac{\partial T}{\partial E},\\
\frac{\partial^{2}I}{\partial A^{2}} &=& \frac{T E}{2\pi A^{2}} - \frac{I}{4 A^{2}} + \frac{E^{2}}{2\pi A^2} \frac{\partial T}{\partial E} .
\eea
The first equation above is obvious. The second one follows immediately
by using the equality (\ref{L=0}). The third one follows in a similar manner,
using (\ref{L=0}) three times and 
$\partial T/\partial A=2\pi \partial^2 I/\partial A\partial E=2\pi\partial/\partial E (\partial I/\partial A)$.
Substituting  the above results in (\ref{Q}) we find in a straightforward manner
the final expression for $Q$,

\be  \label{Qfin}
Q= \frac{I\left(2\pi I\frac{\partial T}{\partial E} + T^{2}\right)}{8 T^{2}}.
\ee
If we have a minimum of (\ref{Taylor}) then we must have $Q>0$, or from (\ref{Qfin}) 

\be  \label{mincond1}
\frac{1}{2} + \frac{\pi I}{T^{2}}\frac{\partial T}{\partial E} \geq 0.
\ee
Using the definition  $\om = 2\pi/T$, and the fact $T = 2\pi \partial I/\partial E$, we find
that the above condition is equivalent to  

\be \label{mincond2}
\frac{\partial \omega}{\partial E}\leq\frac{1}{I},
\ee
or in final form, with a simple geometrical interpretation,

\be \label{mincond3}
\frac{\partial^{2}\left(I^{2}\right)}{\partial E^{2}} \geq 0.
\ee
Namely, the last inequality (\ref{mincond3}) means that $I^2(E)$ is a positive,
monotonically increasing and convex function of $E$ for all $E$ in the range of
the definition of $I(E)$. It could be that it is satisfied only for a 
certain energy range of $E$, in which case we shall say that the
system (potential) has the PR property on the underlying/relevant energy range.

In the next section \ref{PRCY} we shall study many examples of potentials
$V(q)=-Af(q)$, all of them entirely satisfying the PR Conjecture.
Potentials having an escape energy will be always defined (without loss
of generality) in such a way, by an energy shift, that the escape energy 
will be equal to zero. The question arises 
at what value of the parametric kick strength $x=A_1/A_0$ the final
average energy $\langle E_1\rangle\ge 0$ leads to the escape. 
From equation (\ref{aveE1-0}), and reminding that in this case $E_0<0$,
we see immediately that escape takes place if $x\le x_{esc}$, 

\be \label{escape}
x \le x_{esc} = \frac{1}{1 - \frac{T_0 E_0}{\pi I_0}}.
\ee
The potentials with a single minimum satisfying the PR Conjecture
will be treated in the next section \ref{PRCY}, whilst  more complicated case
studies will be presented in section \ref{PRCN}.

\begin{figure} 
\center 
    \includegraphics[width=3.5cm]{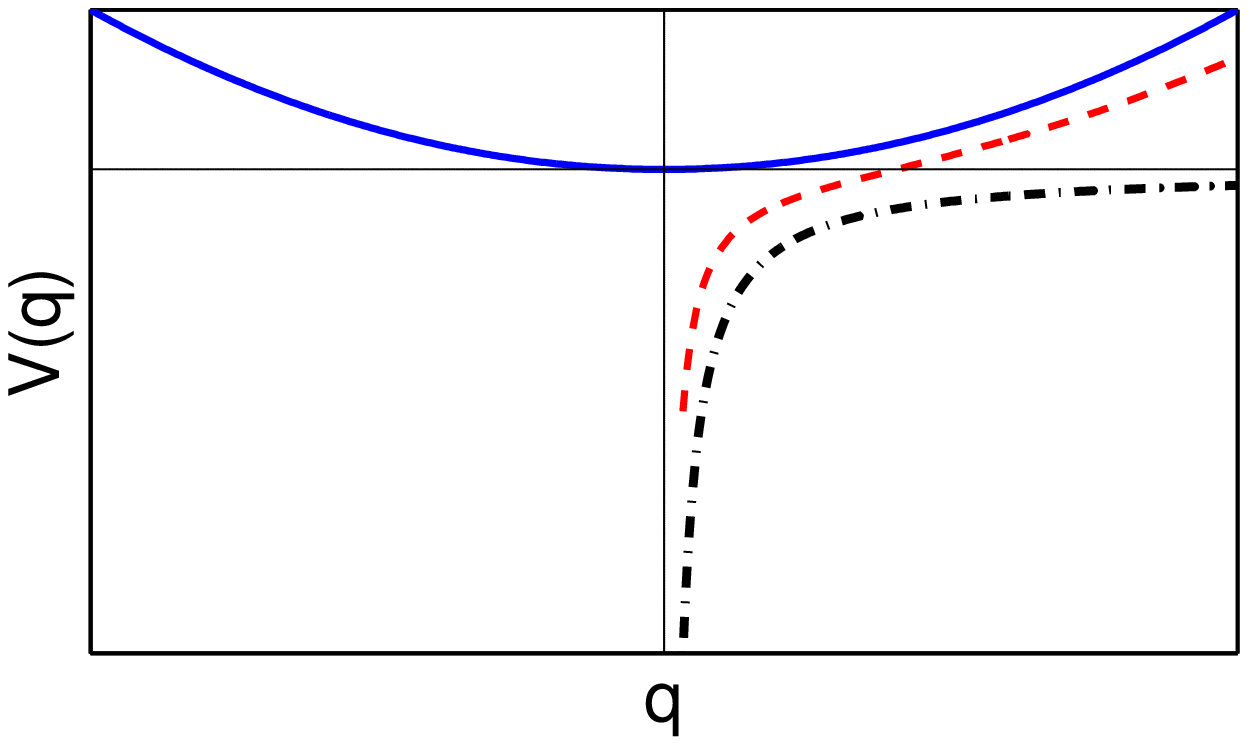} 
    \includegraphics[width=3.5cm]{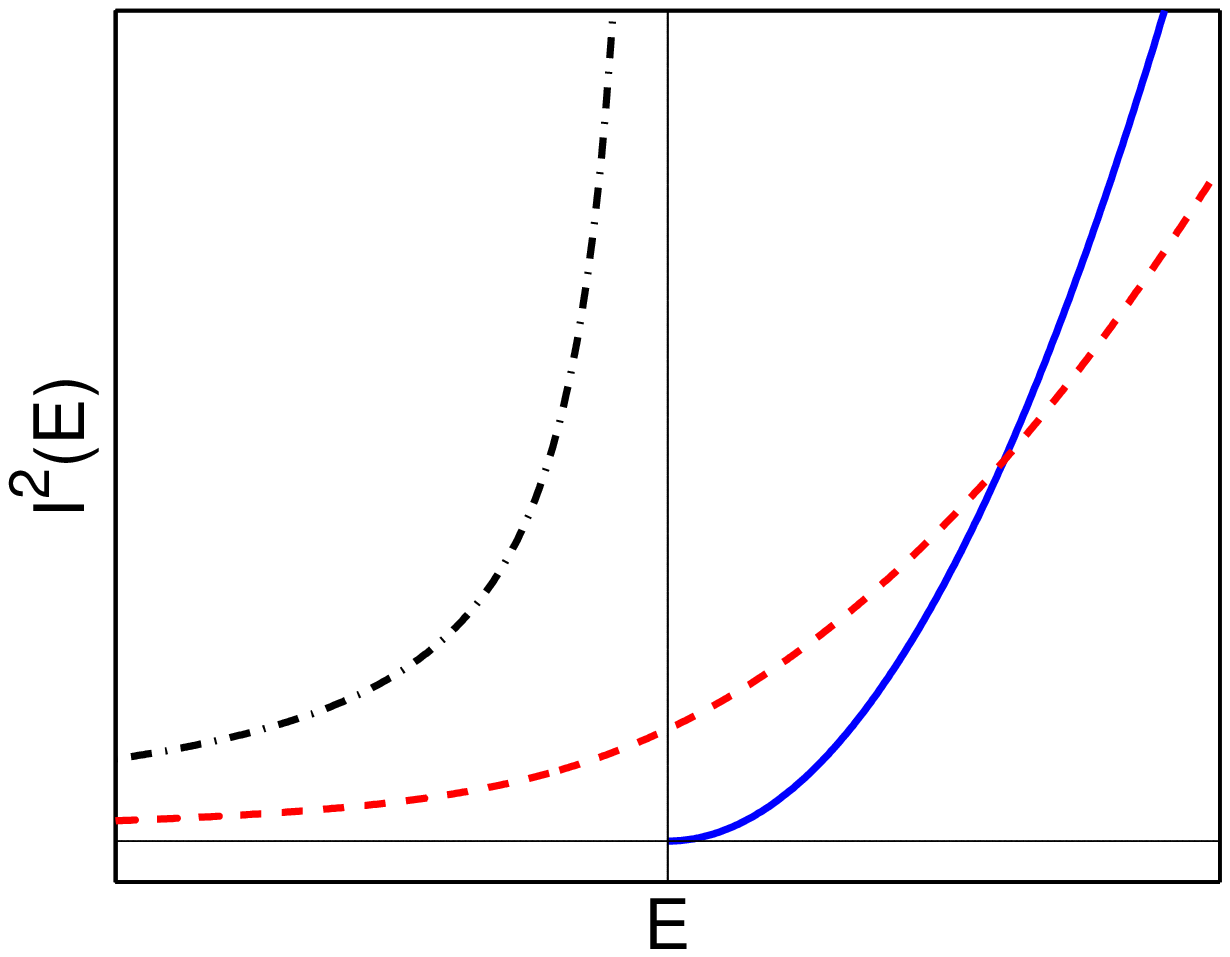} 
    \caption{Sketch of the potentials and the corresponding squared action $I^2(E)$,
which is a positive, monotonically increasing convex function, satisfying the
PR property.}
\label{sketchofpot}
\end{figure}

In figure \ref{sketchofpot} we sketch the behaviour of various potentials and
the associated squared action $I^2(E)$ as a function of the energy $E$,
for which the PR property is satisfied for all $E$.
 
\section{Examples of validity of the PR Conjecture} \label{PRCY}

In this section we give examples of Hamiltonian systems in which the
PR Conjecture (\ref{PRC}) is explicitly and rigorously satisfied
for the entire energy range of the potential $V(q)$.

\subsection{Family of Hamiltonians with homogeneous potential and homogeneous kinetic energy} \label{hompot}

The Hamiltonian is
\begin{equation}\label{homohamiltonian}
H(p,q)=\frac{p^{2n}}{2n} + \frac{Aq^{2m}}{2m},\quad A>0,\; m>0,\; n>0,
\end{equation}
where $m$ and $n$ are positive integers.  The action is

\bea \label{homoaction}
I\left(E\right)&=&\frac{1}{2\pi}\oint  p \; dq  =
\frac{1}{\pi}\int_{-q_{0}}^{q_{0}} \sqrt[2n]{2n\left( E -  \frac{Aq^{2m}}{2m}\right)} \; dq =  
\nonumber \\
 &=& \frac{2^{\frac{2nm+m+n}{2mn}}n^{\frac{1}{2n}}m^{\frac{1}{2m}}}{\pi A^{\frac{1}{2m}}2m}B
\left(\frac{1}{2m},\frac{2n+1}{2n}\right)E^{\frac{n+m}{2mn}},
\eea
where $-q_0$ and $q_0$ are the turning points, and $B$ is the Beta function, 

\begin{equation} \label{betaf}
B \left( x,y \right) = \int_0^{1} t^{x-1}\left( 1-t \right)^{y-1} \; dt. \nonumber
\end{equation}
The period $T_{n,m}$ is related to the frequency $\omega_{n,m}$, as 
$T_{n,m}=\frac{2\pi}{\omega_{n,m}}$. The frequency is

\bea  \label{freqmn}
&&\omega_{n,m}\left( E\right)=\frac{\mathrm{d}E}{\mathrm{d}I}=\left(\frac{\mathrm{d}I}{\mathrm{d}E}\right) ^{-1},  \nonumber \\
&&\frac{dI}{dE}= a(m,n)\;B\left(\frac{1}{2m},\frac{2n+1}{2n}\right)\,E^{\frac{n+m-2mn}{2mn}},\\
&&a(m,n)=\frac{2^{\frac{m+n-2mn}{2mn}}n^{\frac{1}{2n}}m^{\frac{1-4m}{2m}}\left( n+m \right)}{\pi A^{\frac{1}{2m}}}\nonumber
\eea
Now we calculate the average energy after the kick, denoting the kinetic energy
as $K=\frac{p^{2n}}{2n}$, 

\bea
\langle E_1 \rangle &=& \langle K\rangle + \langle V_1 = \rangle \rangle \nonumber \\
&=& \langle K \rangle + \langle V_0 \rangle - \langle V_{0} \rangle + \langle V_1\rangle\nonumber\\
 &=& E_0 + \frac{A_1-A_0}{A_0} \langle V_0 \rangle.
\eea
We use the virial theorem \cite{LL} for this system, rather than equation (\ref{aveE1-0}),
because the kinetic energy $K = \frac{p^{2n}}{2n}$, $n\ge 1$, is more general than quadratic $n=1$
(and this is the only system in this paper in which we have general nonquadratic 
kinetic energy), and obtain 

\begin{equation}
\langle K \rangle =\frac{m}{n} \langle V_0 \rangle \Rightarrow \langle V_0 \rangle =\frac{n}
{m+n} E_0 .
\end{equation}
The ratio of the actions is

\begin{equation}\label{homoactionratio}
\frac{I\left(\langle E_1 \rangle\right)}{I(E_0)}=\frac{\left(1+ \left(x-1\right)\frac{n}{n+m}\right)^{\frac{m+n}{2m n}}}{x^{\frac{1}{2m}}}=F_{n,m}(x).
\end{equation}
The function $F_{n,m}(x)$ has only one minimum at $x=1$ and 
the value of the function is 
equal to $F_{n,m}(1)=1$, as stated by the PR Conjecture.

\subsection{Pendulum}

The Hamiltonian is

\begin{equation}\label{pendulumham}
H(q,p)= \frac{p^{2}}{2} - \Omega^{2} \cos q.
\end{equation}
The action is

\bea
I\left(E\right)=\frac{1}{2 \pi} \oint p \; dq =
\frac{1}{ \pi} \int_{-q_0}^{q_0} \sqrt{2\left(E + \Omega^{2} \cos q\right)} \; dq,
\eea
where $-q_0$ and $+q_0$ are the two turning points in the case of libration (oscillation), defined by $E+\Omega^2\cos q_0=0$.
The action depends on the region in the phase space that we consider. 
There are three regions of energy $E$ and  $q_0$  as follows:

\begin{enumerate}
\item outside the separatrix, $E>\Omega^{2}$, $q_0=\pi$,
\item on the separatrix, $E=\Omega^{2}$, $q_0=\pi$
\item inside the separatrix, $E<\Omega^{2}$, $q_0<\pi$. 
\end{enumerate}

\begin{figure} 
\center 
    \includegraphics[width=3.5cm]{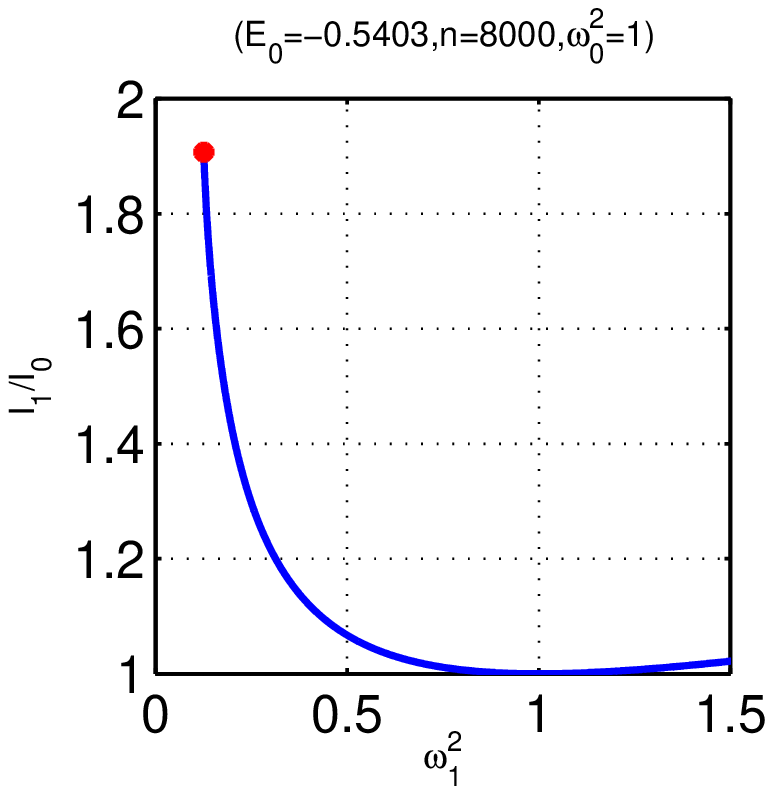} 
    \includegraphics[width=3.5cm]{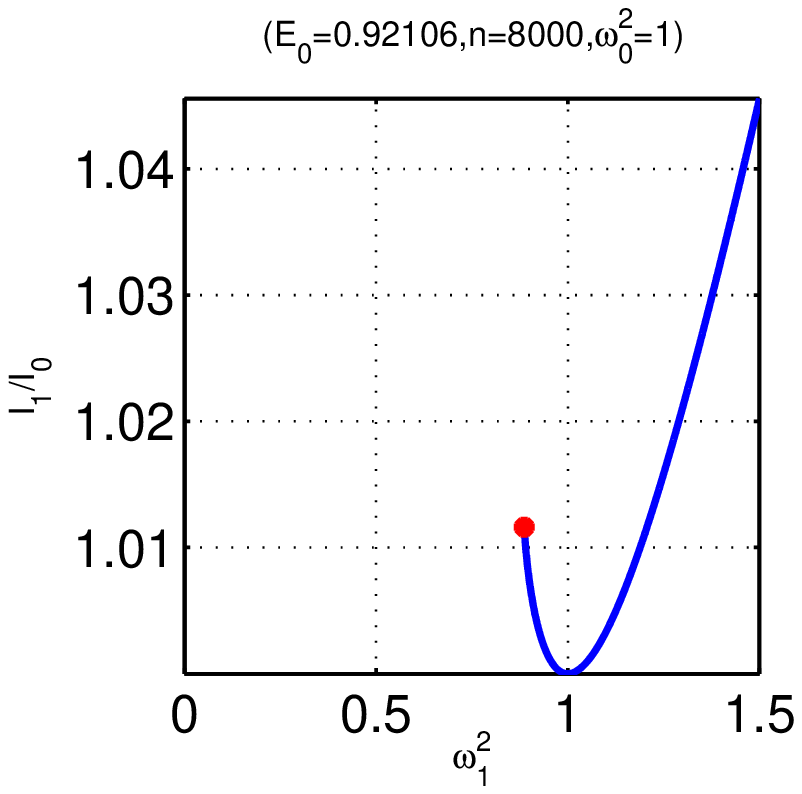} 
    \caption{Ratio of initial and final actions for the pendulum, inside the separatrix, with $E_0$ is the initial energy and $n$ is the number of the points of initial contour. The bullet is the point $(x_{sep},I(x_{sep}))$, which corresponds to the  crossing the separatrix after the kick.}
\label{pendactionrat}
\end{figure}

\noindent
We denote $k=\sqrt{\frac{E+\Omega^{2}}{2\Omega^{2}}}$, and obtain

\begin{equation}
I\left(E\right)=\frac{4}{\pi}\sqrt{2(E+\Omega^{2})}\int_{0}^{\frac{q_0}{2}} \sqrt{1-\frac{\sin^{2}\phi}{k^{2}}} \; d\phi.
\end{equation}
For the first case, $k \ge 1$, we have

\begin{equation}\label{pendulumactionout}
I\left(E\right)=\frac{8\Omega k}{\pi} \rm{E} \left( \frac{1}{k} \right),
\end{equation}
where $\rm{E}$ is complete elliptic integral of the second kind, which is defined as

\be 
{\rm{E}} \left( k \right)=\int_0^{\frac{\pi}{2}} \sqrt{1-k^{2}\sin^{2}\theta} \; d\theta.
\ee
For the second case, $k=1$,  we have, $I\left(E\right)=\frac{8\Omega}{\pi}$.
For the last case, $k \le 1$, we have

\begin{equation}\label{pendulumactionin}
I\left(E\right)=\frac{8\Omega}{\pi} \left[ {\rm{E}} \left(k\right) -(1-k^{2}) {\rm{K}} \left(k\right) \right],
\end{equation}
where $\rm{E}$ is the same complete elliptic integral of the 
second kind and the $\rm{K}$ is elliptic integral of the first kind, 
which is defined as

\begin{displaymath}
{\rm{K}}\left(k\right)=\int_0^{\frac{\pi}{2}} \frac{d\theta}{\sqrt{1-k^{2}\sin^{2}\theta}}.
\end{displaymath}
The period $T(E)$ is related to the frequency $\omega(E)$  as $T=\frac{2\pi}{\omega}$, and the frequency is

\begin{displaymath}
   \omega\left( E\right)=\frac{\mathrm{d}E}{\mathrm{d}I}=\left(\frac{\mathrm{d}I}{\mathrm{d}E}\right) ^{-1}= \frac{\pi \Omega}{2} \times \left\{
     \begin{array}{lr}
\frac{k}{{\rm{K}}\left(\frac{1}{k}\right)}, & \mbox{}  k\in(1,\infty)\\
       \frac{1}{{\rm{K}}\left(k\right)}, & \mbox{}  k\in[0,1] \\
     \end{array}
   \right.
\end{displaymath} 
For the average energy we also have two cases.
For outside the separatrix, $E_0 \geq \Omega_0^{2}$, we get

\bea
\langle E_1 \rangle &= E_0 x + \frac{2\Omega_0^2\; k_0\;{\rm{E}}\left(\frac{1}{k_0}\right)}{{\rm{K}}\left(\frac{1}{k_0}\right)}(1-x),
\eea
whilst for inside the separatrix, $ E_0\le \Omega_0^{2}$, we have

\begin{equation}
\langle E_1 \rangle = E_0 x + 2\Omega_0^2\left[ \frac{\rm{E}(k_0)}{\rm{K}(k_0)} -\left(1-k_0^{2}\right)\right](1-x),
\end{equation}
where $k_0=\sqrt{\frac{E_0^{2} + \Omega_0^{2}}{2\Omega_0^{2}}}$.
We have calculated the average energy $\langle E_1 \rangle$ from the previous equations,
but also checked it numerically. We find that the difference is of the order of
about $10^{-13}$.
In figure \ref{pendactionrat} we show the action ratio for the pendulum
in the libration regime.

\begin{figure} 
  \center
    \includegraphics[width=4.5cm]{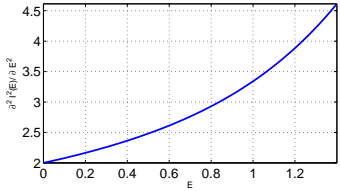}
\caption{The pendulum: The second derivative $d^2I^{2}(E)/dE^2$, with 
$\omega=1$. }
\label{I2Epend}
\end{figure}

\noindent
For the case of libration the action $I(E)$ as a power series 
of $\epsilon = E + \Omega^{2}$ reads as

\bea
I(E) &= \frac{\epsilon}{4194304 \Omega^{11}} ( 1323 \; \epsilon^{5} + 3920 \Omega^{2}\; \epsilon^{4} + 12800 \Omega^{4}\; \epsilon^{3} \nonumber  \\
&+ 49152 \Omega^{6}\; \epsilon^{2} + 262144 \Omega^{8}\; \epsilon  + 4194304 \Omega^{10}  + h.o.t. ). \nonumber \\
\eea
and the function $I^2(E)$ is plotted in  figure \ref{I2Epend}. The PR property
is satisfied.

\subsection{Radial Kepler problem with zero angular momentum} \label{Keplerzero}

The Hamiltonian is

\begin{equation}\label{keplerpotential}
H(r,p)= \frac{p^{2}}{2} - \frac{A}{r},\quad A>0.
\end{equation}
The action is

\bea \label{kepleraction}
I\left(E\right)&=&\frac{1}{2 \pi} \oint p \; dr = \nonumber \\
&=& \frac{1}{\pi} \int_{r_{1}}^{r_{2}} \sqrt{2\left(E + \frac{A}{r}\right)} \; dr = \frac{A}{\sqrt{2\vert E\vert}},
\eea
where $r_1$ and $r_2$ are the two turning points. In fact, $r_1=0$.
The period $T$ is related to the frequency $\omega$, as $T=\frac{2\pi}{\omega}$,

\begin{equation}
\omega\left( E\right)=\frac{\mathrm{d}E}{\mathrm{d}I}=\left(\frac{\mathrm{d}I}{\mathrm{d}E}\right) ^{-1}=\frac{2\sqrt{2}\;\vert E \vert ^{\frac{3}{2}}}{A}.
\end{equation}
The average energy is

\begin{equation}
\langle E_1 \rangle = \langle K \rangle + \langle V_1 \rangle = E_0 + \frac{A_1 - A_0}{A_0} \langle V_0
\rangle. \nonumber
\end{equation}
We can use the virial theorem \cite{LL} to calculate the average potential,
 
\begin{equation}
\langle E_0 \rangle = \frac{\langle V_0 \rangle}{2} = E_0.
\end{equation}
Finally, we get the ratio of the actions $ I_1 $ and $ I_0 $

\begin{equation}\label{kepleractionratio}
\frac{I\left(\langle E_1 \rangle\right)}{I(E_0)} = \frac{x}{\sqrt{2x - 1}}=F(x),
\end{equation}
where $ x = A_1/A_0$.
For this Kepler problem we have positive average energy $\langle E_1\rangle$ if $A_1 \le A_0/2$.
Thus the ratio of the actions $I_1 \geq I_0$, for no-escape orbits, as the function 
$F(x)$ has the minimum at $x=1$ and the value of the function is $F(1)=1$ in accordance
with the PR Conjecture.

\subsection{Radial Kepler problem with nonzero angular momentum}

\begin{figure} 
  \center
    \includegraphics[width=3.5cm]{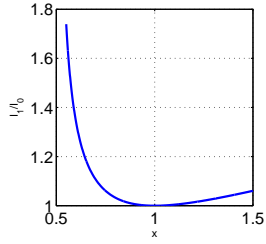}
  \caption{Action ratio for the Kepler potential with nonzero angular momentum, 
with $A_0=1$ and $c=\frac{1}{2}$. The $x$ for escape orbit is $x\le x_{esc}=\frac{1}{3}$, 
with action ratio $I_1/I_0 \rightarrow \infty$ as $x\rightarrow x_{esc}$.}
\label{actratKepler}
\end{figure}

The Hamiltonian with the angular momentum $M$ is
\begin{equation}\label{keplereffectivepotential}
H(r,p)= \frac{p^{2}}{2} - A \left(\frac{a}{r} -  \frac{M^{2}}{2r^{2}}\right),\quad A>0,\; a>0,\;M^{2}>0.
\end{equation}
The action is
\bea \label{keplereffectiveaction}
I\left(E\right)&=&\frac{1}{2 \pi} \oint  p \; dq =\frac{1}{\pi} \int_{r_{1}}^{r_{2}} 
\sqrt{2\left(E + A\left(\frac{a}{r} - \frac{M^{2}}{2r^{2}}\right)\right)} \; dr \Rightarrow \nonumber \\
I\left(E\right)&=& \frac{A \; a}{\sqrt{2\vert E\vert}} - \sqrt{A}\; M.
\eea
Here $r_1$ and $r_2$ are the two turning points. The frequency is

\begin{equation}
\omega\left( E\right)=\frac{\mathrm{d}E}{\mathrm{d}I}=\left(\frac{\mathrm{d}I}{\mathrm{d}E}\right) ^{-1}=
\frac{2\sqrt{2} \; \vert E \vert ^{\frac{3}{2}}}{A\; a}.
\end{equation}
The average energy is, using equation (\ref{aveE1-0}), 

\begin{equation}
\langle E_1 \rangle = E_0 x + \frac{\sqrt{2}\;\vert E_0 \vert ^{\frac{3}{2}}}{A_0 \;a}
\left(\frac{A_0 \; a}{\sqrt{2\;\vert E_0 \vert}} - \sqrt{A_0}\;M\right)(1-x),
\end{equation}
where $x=A_1/A_0$.
\\
We have positive average energy after the kick, if 
\begin{displaymath}
x \leq x_{esc}= \frac{A_0 \; a - 
\sqrt{2 A_0}\;M\vert E_0 \vert^{\frac{1}{2}}}{2 A_0 \; a - \sqrt{2 A_0}\;M\vert E_0 \vert^{\frac{1}{2}}}.
\end{displaymath}
For the action ratio we have finally
\begin{equation} 
\frac{I\left(\langle E_1 \rangle\right)}{I\left(E_0\right)} =
\frac{x \sqrt{\frac{1}{2x - 1 + (1-x)c}} - c \sqrt{x}}{1 - c}=F(x),
\end{equation}
where $c=\frac{M\sqrt{2 \vert E_0 \vert}}{a \sqrt{A_0}}$. Since the minimal energy $E_{min} \le 0$ 
is determined by the condition $I(E_{min})=0$, we find from (\ref{keplereffectiveaction})
$|E_{min}| = a^2A/(2M^2)$, and thus due to $|E_0| \le E_{min}$ we have $c \le 1$.
If $M=0$ and consequently $c=0$, 
we recover the result for the Kepler problem with zero angular momentum of the previous
subsection \ref{Keplerzero}. The derivative of $F(x)$ is

\bea
\frac{\partial F(x)}{\partial x} = \frac{-\frac{x}{2}(2-c)\left(\frac{1}{B(x)}\right)^{\frac{3}{2}} + 
\left(\frac{1}{B(x)}\right)^{\frac{1}{2}} - \frac{c}{2\sqrt{x}}}{1-c},
\eea
where $B(x) = (2-c)x + c-1$.
There is only one root for $\frac{\partial F(x)}{\partial x} = 0$, namely precisely $x = 1$.
For the second derivative of $F(x)$ we find

\bea
\frac{\partial^{2} F(x)}{\partial x^{2}} = 
\frac{\frac{3}{4}(c-2)^{2}x\left(\frac{1}{B(x)}\right)^{\frac{5}{2}} +
 (c-2)\left(\frac{1}{B(x)}\right)^{\frac{3}{2}} + \frac{c}{4 x^{3/2}}}{1-c},
\eea
and at $x = 1$ we have

\bea
\frac{\partial^{2} F(x)}{\partial x^{2}}\vert_{x=1} = 1 - \frac{3}{4}c > 0
\eea
for all $c \le 1$. Thus there is only one minimum for the function $F(x)$ at $x=1$, 
with $F(1)=1$, exactly in accordance with the PR Conjecture. In figure
\ref{actratKepler} we show the action ratio as a function of $x$ for $c=1/2$.

\subsection{Morse potential} \label{Morse}

\begin{figure}   
  \center
    \includegraphics[width=3.5cm]{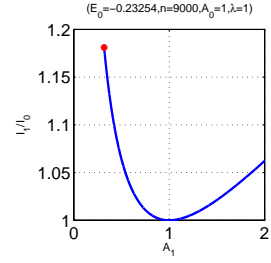}
  \caption{Action ratio for the Morse potential, with $\lambda=1$. 
$A_0=1$ and $n$ is the number of points in initial contour. 
$x$ for escape orbits is $x \le x_{esc}=0.31803$,
with the action ratio $I_1/I_0=1.18103$, marked by a bullet.}
\label{morsefig}
\end{figure}

The Hamiltonian is

\begin{equation}\label{morsepotential}
H(q,p)= \frac{p^{2}}{2} + A(e^{-2 \lambda r} - e^{- \lambda r}),\quad A>0,\quad \lambda>0.
\end{equation}
The minimum of the potential is equal to $-A/4$. The action is

\bea\label{morseaction}
I\left(E\right) = \frac{1}{2 \pi} \oint  p \; dr 
= \frac{1}{\pi} \int_{r_1}^{r_2} \sqrt{2\left(E - A(e^{-2 \lambda r} 
- e^{- \lambda r})\right)} \; dr, \nonumber
\eea
where $r_1$ and $r_2$ are the two turning points, and we get

\begin{equation}
I\left(E\right) = -\frac{\sqrt{2\vert E \vert}}{\lambda} + \frac{\sqrt{A}}{\lambda \sqrt{2}}.
\end{equation}
The frequency $\omega$ is
\begin{equation}
\omega\left( E \right)=\frac{\mathrm{d}E}{\mathrm{d}I}=
\left(\frac{\mathrm{d}I}{\mathrm{d}E}\right) ^{-1}=\lambda \sqrt{2 \vert E \vert}.
\end{equation}
For the average energy, we have
\begin{equation}
\langle E_1 \rangle = \langle K \rangle + \langle V_1 \rangle = E_0 + \frac{A_1-A_0}{A_0}\langle V_0 \rangle. \nonumber
\end{equation}
In contradistinction to some other systems, here we can find the $r$ as a function of time $t$ in
an explicit closed form

\bea \label{morseactionratio}
&&\frac{dr}{dt}=\sqrt{2(E-A(e^{-\lambda r} - e^{-2\lambda r}))}\Rightarrow \nonumber \\
&& t= \int \frac{dr}{\sqrt{2(E-A(e^{-\lambda r} - e^{-2\lambda r}))}}.
\eea
Using $u=e^{-\lambda r}$ and after some calculations we obtain

\be
t = -\frac{\arcsin\left( \frac{2E+Au}{u\sqrt{4AE+A^{2}}}\right) }{\lambda\sqrt{2\vert E \vert}},
\ee
or,

\begin{equation} \label{Morseroft}
r(t) = -\frac{1}{\lambda}\ln \left[ \frac{2\vert E \vert}{\sqrt{4EA+A^{2}}\; 
\sin \left(\sqrt{2\vert E \vert}\;\lambda t\right) +A}\right].
\end{equation}
Finally the average energy is, either using the equation 
(\ref{aveE1-0}) or doing the direct averaging over time  using (\ref{Morseroft}),

\be \label{Morseaverageenergy}
\langle E_1 \rangle = E_0 + \frac{\sqrt{A_0 \vert E_0 \vert}}{2}(1-x),
\ee
where  $x=A_1/A_0$.
We have positive average energy after the kick, at 
$x \leq x_{esc}= 1 - 2\vert E_0 \vert/A_0$.
For the action ratio we have

\begin{equation}
\frac{I\left(\langle E_1 \rangle\right)}{I\left(E_0\right)} = 
\frac{ \sqrt{x} -2 \sqrt{ \vert \frac{E_0}{A_0} + \frac{1}{2}\sqrt{\frac{\vert E_0 \vert}{A_0}}\; (1 - x)\vert}}
{1 - 2\sqrt{\frac{\vert E_0 \vert}{A_0}}}=F(x).
\end{equation}
We define $E_0 = -A_0/4 + \epsilon$, $0\leq \epsilon \leq A_0/4$, and 
$c=\epsilon/A_0$, so that $0\le c \le 1/4$, and simplify

\bea
 F(x) = \frac{I\left(\langle E_1 \rangle\right)}{I\left(E_0 \right)} = \frac{\sqrt{x} -2 \sqrt{\vert -c + \frac{1}{4} + \frac{\sqrt{-c + \frac{1}{4}}}{2}\; (1 - x)\vert}}{1 - 2\sqrt{\vert \frac{1}{4}-c \vert}}.
\eea
The function $F(x)$ has a single minimum at $x=1$ and $F(1)=1$, in agreement with the PR Conjecture. In figure \ref{morsefig} we plot the action ratio
as a function of $A_1$ at $\la=1$ and $A_0=1$

\subsection{P\"oschl-Teller I potential} \label{PTI}

The Hamiltonian is

\begin{equation}  \label{pt1hamiltonian} 
H(q,p) = \frac{p^{2}}{2} + \frac{A}{\cos^{2} \left( \lambda q \right)},\quad \lambda>0,\; A>0.
\end{equation}
The action is

\bea \label{pt1action}
I\left(E\right) &= \frac{1}{2 \pi} \oint  p \; dq 
= \frac{1}{\pi} \int_{q_1}^{q_2} \sqrt{2\left(E - \frac{A}{\cos^{2} \lambda q}\right)} \; dq, \nonumber
\eea
where $q_1$ and $q_2$ are the two turning points.
Using  transformation, $ y=\tan \left(\lambda q\right) $ and after calculation, we get finally,

\begin{equation}
I\left(E\right) = \frac{\sqrt{2}}{\lambda} (\sqrt{E} - \sqrt{A}).
\end{equation}
The frequency is

\begin{equation}
\omega\left( E\right)=\frac{\mathrm{d}E}{\mathrm{d}I}=
\left(\frac{\mathrm{d}I}{\mathrm{d}E}\right) ^{-1}=\lambda\sqrt{2E}.
\end{equation}
Now we calculate the final average energy either using (\ref{aveE1-0}) or by
direct averaging 

\be
\langle E_1 \rangle = E_0 + \frac{A_1 - A_0}{A_0} \langle V_0 \rangle. \nonumber
\ee
where 

\begin{equation}
\langle V_0 \rangle = \frac{1}{T} \int_{0}^{T} \frac{A_0}{\cos^{2} \left( \lambda q(t) \right)} \; dt.
\end{equation}
Here again we can express $q$ as function of time $ t $, namely, after some straightforward
calculation, we obtain

\begin{equation}
q(t) = \frac{1}{\lambda} \arcsin \left( \sqrt{\frac{E-A}{E}} \sin \left( \lambda t\sqrt{2E} \right)\right).
\end{equation}
Further, using the transformation $ y= \lambda \sqrt{2E_0} t $, we find

\bea 
\langle V_0 \rangle &=& \frac{1}{T} \int_{0}^{T} \frac{A_0}{\cos^{2} \left( \lambda q(t) \right)} \; dt
\nonumber\\
&=& \frac{A_0}{\lambda T \sqrt{2E_0}} \int_{0}^{2\pi} \frac{dy}{1-\frac{E_0 - A_0}{E_0}
\sin^{2} y} \Rightarrow \nonumber \\
\langle V_0 \rangle &=& \sqrt{A_0E_0}.
\eea
The average energy, with $x=A_1/A_0$, is thus 

\be 
\langle E_1 \rangle = E_0 +(x-1) \sqrt{E_0A_0}.
\ee
The minimum energy is at $E_0=A_0$, so we write $E_0=A_0+\epsilon$, where $\epsilon\geq 0$. Then, setting
$c=\epsilon/A_0$,   we obtain the action ratio $F(x)$ as

\begin{equation}
\frac{I\left(\langle E_1 \rangle\right)}{I\left(E_0\right)}=F(x)=
\frac{\sqrt{ c+1 +\sqrt{ c+1 }(x-1)}-\sqrt{x}}{\sqrt{ c+1 }-1}.
\end{equation}
This function $F(x)$ has a single minimum at $x=1$ and $F(1)=1$, in agreement with the PR Conjecture.

\subsection{P\"oschl-Teller II potential} \label{PTII}

The Hamiltonian is

\begin{equation}\label{pt2hamiltonian}
H(q,p) = \frac{p^{2}}{2} - \frac{A}{\cosh^{2} \left(\lambda q\right)},\quad \lambda>0,\; A>0.
\end{equation}
The action  is

\bea \label{pt2action}
I\left(E\right) &= \frac{1}{2 \pi} \oint  p \; dq =
\frac{\sqrt{2}}{\lambda}\left(\sqrt{A}-\sqrt{\vert E \vert}\right).
\eea
The frequency is

\begin{equation}
\omega\left( E\right)=\frac{\mathrm{d}E}{\mathrm{d}I}=\left(\frac{\mathrm{d}I}{\mathrm{d}E}\right) ^{-1}
=\lambda\sqrt{2 \vert E \vert}.
\end{equation}
The final average energy is
\bea
\langle E_1 \rangle &= \langle K \rangle + \langle V_1 \rangle = \langle K \rangle + \langle V_0
\rangle - \langle V_0 \rangle + \langle V_1 \rangle \Leftrightarrow \nonumber \\
\langle E_1 \rangle &= E_0 + \frac{A_1 - A_0}{A_0} \langle V_0 \rangle, \nonumber
\eea
where we need

\begin{equation}  \label{aveVPTII}
\langle V_0 \rangle = \frac{1}{T} \int_{0}^{T} \frac{A_0}{\cosh^{2} \left(\lambda q(t)\right)} \; dt.
\end{equation}
In this case again we can find the explicit solution of $q(t)$ in a closed form, namely

\begin{equation}
q(t) = \frac{1}{\lambda} \rm{arcsinh} \left( \sqrt{\frac{E+A}{\vert E \vert}} 
\sin \left( \lambda t \sqrt{2\vert E \vert}\right)\right).
\end{equation}
The average potential is

\bea
\langle V_0 \rangle &=& -\frac{1}{T} \int_{0}^{T} \frac{A_0}{\cosh^{2}\left( \lambda q(t)\right)} \; dt \nonumber \\
&=& -\frac{A_0}{\lambda T \sqrt{2\vert E_0 \vert}} \int_{0}^{2\pi} 
\frac{dy}{1+\frac{E_0 + A_0}{\vert E_0 \vert}\sinh^{2} y} \Rightarrow \nonumber \\
\langle V_0 \rangle &=& -\sqrt{A_0 \vert E_0 \vert},
\eea
The minimum of the potential is $-A_0$, so we introduce $\epsilon$, such that $E_0=-A_0 +\epsilon$. 
Further, by defining $c=\epsilon/A_0$, we can calculate the
ratio of the actions after the kick, as follows

\begin{equation}
\frac{I\left(\langle E_1 \rangle\right)}{I\left(E_0 \right)}=
F(x)=\frac{\sqrt{\vert c-1 +\sqrt{\vert c-1 \vert}(x-1)\vert} -\sqrt{x}}{\sqrt{\vert c-1 \vert}-1}.
\end{equation}
We have positive average energy after the kick if 
$x\le x_{esc} =  1-\sqrt{\vert E_0 \vert/A_0}$.
For $x\ge x_{esc}$ the function $F(x)$ has a single minimum at $x=1$ and its value is $F(1)=1$, 
in complete agreement with the PR Conjecture.

\subsection{Cosh potential} \label{Coshpotential}

\begin{figure}   
  \center
    \includegraphics[width=4.5cm]{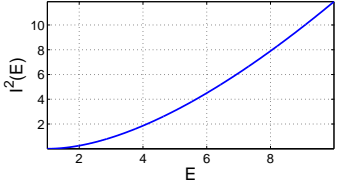}
  \caption{Cosh potential:  $I^2(E)$ as a
function of $E$, for $A=1$ and $\la=1$, showing that it is
a convex function, satisfying the PR property. }
\label{actioncosh}
\end{figure}

The Hamiltonian is

\begin{equation}\label{coshhamiltonian}
H(q,p) = \frac{p^{2}}{2} +  A \cosh \lambda q,\quad A>0.
\end{equation}
The action is

\bea
I(E) &=& -\frac{i\,4\sqrt{2(E-A)}}{\pi \lambda} \nonumber\\
&&\quad\,{\rm E} \left(\frac{i}{2} \log
\left( \frac{E+\sqrt{E^2+A^2}}{A} \right), \frac{2A}{A-E} \right),
\eea
where ${\rm E}$ is the elliptic integral of the second kind, with $k^2=2A/(A-E)$,
and the definition

\be 
{\rm E}(\phi,k) = \int_0^\phi dt \sqrt{1 - k^2\sin^2 t}.
\ee
For large energies $E \gg A$ we have the asymptotics  $I(E) \approx \sqrt{E}\log E$, and 
the second derivative of $I^{2}$ is approximately

\bea
\frac{\partial ^{2} I^{2}(E)}{\partial E^{2}} \propto
\frac{2\log E}{E},
\eea
which means again that the PR Conjecture is satisfied. In figure \ref{actioncosh}
we plot the function $I^2(E)$ as a function of $E$ for the case
$A=0$ and $\la=0$.

\section{Studies of breaking the PR Conjecture} \label{PRCN}

In this section we shall investigate the potentials where the PR Conjecture
is violated, either because of the nonanalyticity (nonsmoothness) or due to 
the closeness to a stationary point (separatrix in the phase space).
Since the PR Conjecture in full generality is not true, we will better speak
of the PR property rather than PR Conjecture. We shall show the breaking of
the PR property in case of $C^0$ potential in the subsection \ref{box}, and
then demonstrate in the two next subsections that PR property is restored
if the first derivative is continuous, and even more so if the second
derivative of the potential $V(q)$ is continuous. After that we
shall study what happens if the potential is not a single
well potential, but has other stationary points, implying the existence
of a separatrix in the phase space. In those cases the PR property can
be broken, typically  for a certain energy range around the
separatrix.

\subsection{Harmonic oscillator in a box} \label{box}

The first example of a potential (system) which violates the PR Conjecture
is a nonanalytic potential, namely the harmonic oscillator in a box.
The Hamiltonian is
\begin{equation}\label{boxhamiltonian}
H(q,p) = \frac{p^{2}}{2} + V(q), 
\end{equation}
where

\begin{displaymath}
   V(q)=\left\{
     \begin{array}{lr}
       \frac{\omega^{2} q^{2}}{2}, & \mbox{}  q\in[-q_{0},q_{0}] \\
	   q = \infty , & \mbox{}  |q| \ge q_0
     \end{array}
   \right\},
\end{displaymath}
and the kick parameter is $A=-\om^2$.
The potential is continuous, but has discontinuous first derivative at
$q=q_0$, so it is $C^0$ only. For the energy $E$ smaller than 
$E_c = \om^2q_0^2/2$ the system behaves just as the ordinary linear
oscillator treated in subsection \ref{hompot}, equation (\ref{homohamiltonian}), 
and $d^2I^2/dE^2=1/\om^2\ge 0$, 
whilst at $E\ge E_c$ the box potential plays a role. For a pure box
potential without the harmonic oscillator we have $I\propto \sqrt{E}$
and consequently $d^2I^2/dE^2=0$. As we will see, for the combined 
potential (\ref{boxhamiltonian}) at $E\ge E_c$ we find that 
$d^2I^2/dE^2<0$.  The action at $E\ge E_c$  is 

\bea  \label{boxaction}
I\left(E\right) &= \frac{1}{2 \pi} \oint p \; dq
= \frac{1}{\pi} \int_{-q_0}^{q_0} \sqrt{2\left(E + \frac{1}{2} \omega^{2} q^{2}\right)} \; dq.
\nonumber \\
I\left(E\right)&=\frac{2 E}{\pi \omega}\left({\rm{arcsin}} \left( \frac{q_{0}\omega}{\sqrt{2E}} \right) + \frac{q_{0} \omega}{\sqrt{2E}}\sqrt{1 - \frac{q_{0}^{2} \omega^{2}}{2E}} \;\right).
\eea
The action in series expansion is

\bea
I(E) = \frac{4 E x_{0}}{\pi \omega} \left( 1 - \frac{x_{0}^{2}}{6} + \frac{3 x_{0}^{4}}{20} - \frac{x_{0}^{6}}{112} + h.o.t. \right)
\eea
where $x_0 = \frac{q_{0}\omega}{\sqrt{2 E}}$. Finally, we have
\begin{equation}
\frac{\partial ^{2} I^{2}(E)}{\partial E^{2}} =  - \frac{16 x_{0}^{6}}{45 \pi^{2} \omega^{2}}\left( 2 + \frac{18}{7}x_{0}^{2} + h.o.t. \right) \leq 0
\end{equation}
Thus we see that due to the harmonic potential inside the box $d^2I^2/dE^2$
is no longer zero, but negative.
Therefore, the curve $I^2(E)$ is not convex for the energy range
$E \ge E_c=\om^2q_0^2/2$ and implies that $Q$ defined in
equation (\ref{Taylor}) and given in the closed form in (\ref{Qfin}),
is negative and therefore $F(x)$ has
a maximum $F(1)=1$ at $x=1$ instead of a minimum, which is a violation of
the PR property at energy range $E\ge E_c$.

\subsection{Quadratic-linear potential}  \label{qlpot}

\begin{figure} 
  \center
    \includegraphics[width=4.5cm]{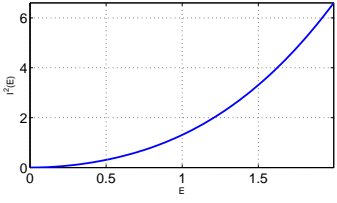}
  \caption{Quadratic-linear potential. 
The $I^{2}(E)$ with $A=1$, $\omega = 1$ and $q_0=1$.}
 \label{I2Eql}
\end{figure}

The previous nonanalytic potential does not possess the PR property,
due to nonanalyticity. In fac, $V(q)$ is only a continuous function,
the first derivative is discontinuous. Therefore, let us look at an example
where the first derivative is smooth, whilst the  second is not.
Thus we construct a potential which is quadratic (harmonic) up
to $|q| \le q_0$, and linear outside the interval $(-q_0,q_0)$, such
that the first derivative $dV/dq$ at $|q|=q_0$ is continuous.
The Hamiltonian is

\begin{equation}\label{qlhamiltonian}
H(q,p) = \frac{p^{2}}{2} - A \; f(q),\quad A>0.
\end{equation}
where

\begin{displaymath}
   f(q)=\left\{
     \begin{array}{lr}
       -\frac{\omega^{2} q^{2}}{2}, & \mbox{}  q\in[-q_{0},q_{0}] \\
- \om^2 q_0|q| + \frac{\om^2q_0^2}{2}, 
& \mbox{}  q \not\in [-q_{0},q_{0}].
     \end{array}
   \right.
\end{displaymath}
In figure \ref{I2Eql} we show 
the plot of the $I^{2}(E)$, obtained numerically, 
and it is obvious that it is a convex function for all $E$,
meaning that the PR property is satisfied. We might conjecture that
the $C^1$ smoothness of a single minimum potential $V(q)=-Af(q)$ is enough
for the PR property to hold.

\subsection{Quadratic-quartic potential} \label{qqpot}

\begin{figure} 
  \center
    \includegraphics[width=4.5cm]{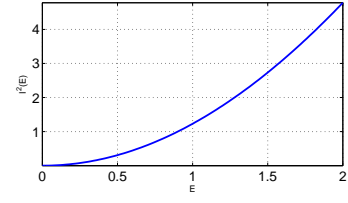}
  \caption{Quadratic-quartic potential: $I^{2}(E)$ with $A=1$, $\omega = 1$
and $q_0=1$.}
\label{I2Eqq}
\end{figure}

Let us increase the degree of smoothness and consider the $V(q)$ functions
of class $C^2$, i.e. having a continuous second derivative only.
Of course, we expect the PR property to hold, and this is indeed
observed.  The Hamiltonian is 

\begin{equation}\label{qqhamiltonian}
H(q,p) = \frac{p^{2}}{2} - A \; f(q),\quad A>0,
\end{equation}
where
\begin{displaymath}
   f(q)=\left\{
     \begin{array}{lr}
       -\frac{\omega^{2} q^{2}}{2}, & \mbox{}  q\in[-q_{0},q_{0}] \\
       -\frac{\omega^{2}(q + 2 q_{0}\; {\rm sign}(q))^{4}}{108\;q_{0}^{2}} +
 \frac{\omega^{2} q_{0}^2}{4} , & \mbox{} 
q \not\in [-q_{0},q_{0}].
     \end{array}
   \right.
\end{displaymath}
We have calculated $I(E)$ numerically, and show the plot of 
the function $I^{2}(E)$ in figure \ref{I2Eqq} and it is obvious that 
it is a convex function of $E$ for all $E$. Thus, the PR property is satisfied.

\subsection{Sextic potential} \label{sexticpot}

\begin{figure} 
  \center
    \includegraphics[width=4.5cm]{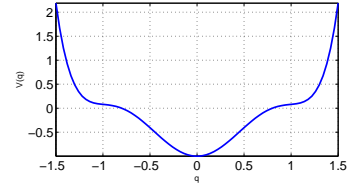}
  \caption{Sextic potential, with $A=1$, $c=0.001$, for $-1.5 \leq q \leq 1.5$.}
\label{sexpotVq}
\end{figure}

The Hamiltonian is

\begin{equation} \label{qlhamiltonian}
H(q,p) = \frac{p^{2}}{2} - A \; f(q),\quad A>0,
\end{equation}
where the potential $V(q)=-Af(q)$ is given by

\begin{equation}
f(q) = -(cq + (q-1)^{3})(cq + (q+1)^{3}), \; c\ge 0.
\end{equation}
This model cannot be analyzed analytically.
If $c=0$, there are two stationary points at $q=\pm 1$, namely
inflection points, at the energy (potential level) $E=V=0$. 
By numerical techniques one can convince
himself that the $dI^2/dE^2$ can be negative if the energy
is close to the value of the stationary points. This is
still true if $c$ is nonzero, but small. We choose $c=0.001$.
The potential is plotted in figure \ref{sexpotVq}, and it 
has only one minimum.

\begin{figure} 
\center
\includegraphics[width=3.5cm]{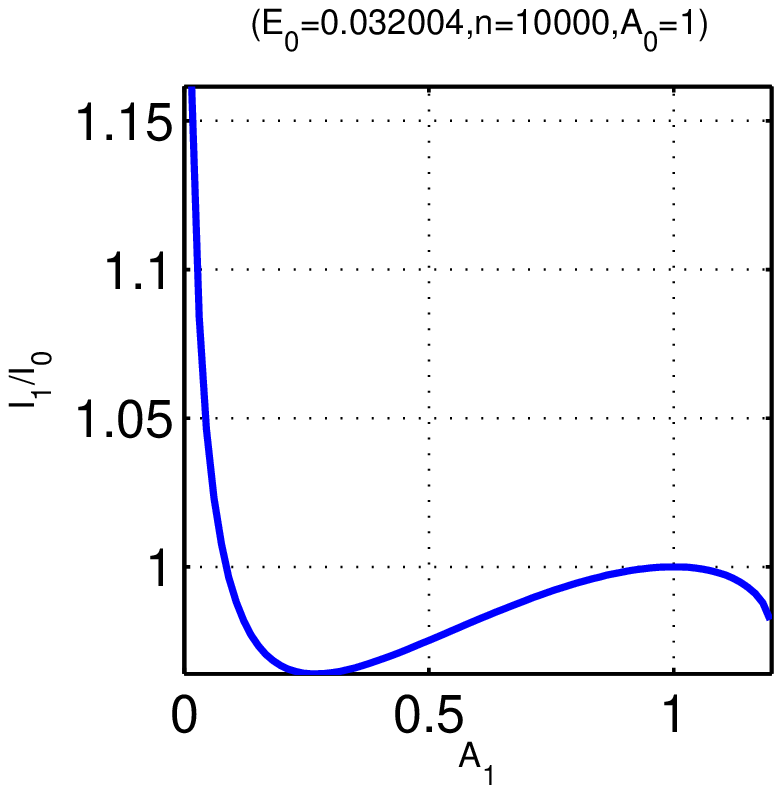}
\includegraphics[width=3.5cm]{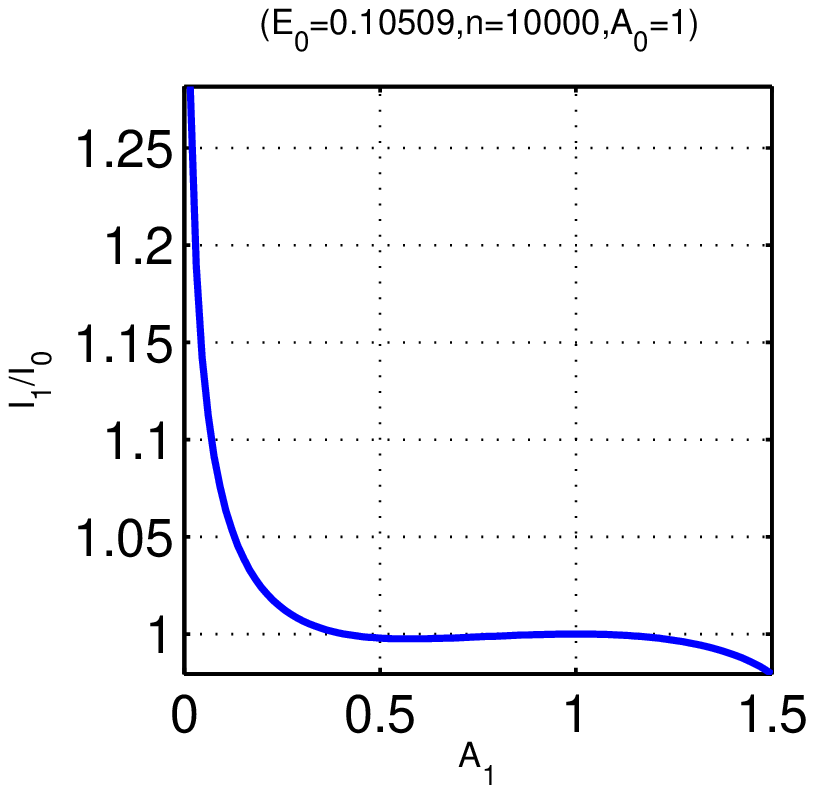}
\includegraphics[width=3.5cm]{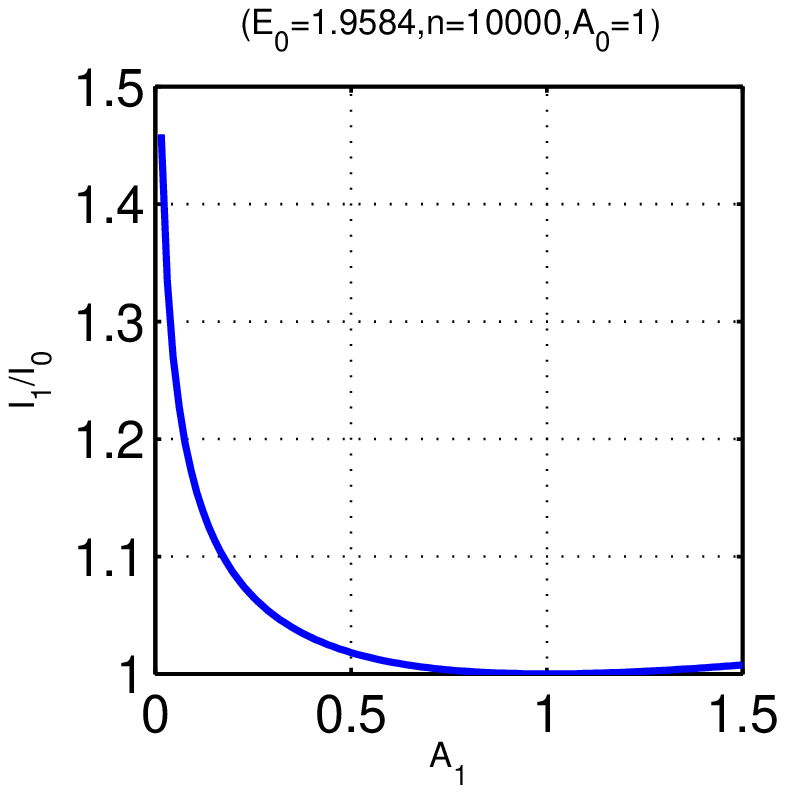}
\caption{Sextic potential. The action ratio $I_{1}/I_{0}$ with $A_{0}=1$, $c=0.001$ and initial energy $E_0$. $n$ is  the number of the points in the initial ensemble. Clockwise: For (a) we find a maximum, (b)  inflection 
point (or very shallow maximum) and in (c) we have a minimum, at $x=A_1/A_0=1$.
Only in the last case (c) PR property is satisfied.}
\label{sexpotF}
\end{figure}

Now let us calculate directly the action ratio $I_1/I_0 =F(x)$ for the
parameters $A_0=1$ as a function of $x=A_1$. We observe three
cases in figure \ref{sexpotF}. 
At high energy $E=1.9584$ there is a minimum of $F(x)$ at
$x=1$, which means that the PR property is satisfied. At small
energy $E=0.032004$ we have a maximum at $x=1$, meaning the violation of
the PR property, whilst at some intermediate energy $E=0.10509$ we
have a very flat maximum, very close to an inflection point.

\subsection{Quartic double well potential} \label{qdwpot}

\begin{figure} 
  \center
    \includegraphics[width=4.5cm]{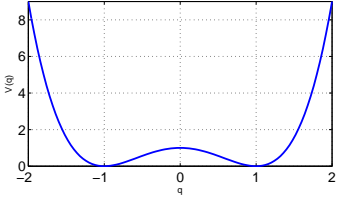}
  \caption{Quartic double well potential, with $A=1$, for $-2 \leq q \leq 2$.}
\label{qdwpotfig}
\end{figure}

\begin{figure} 
\center
\includegraphics[width=3.5cm]{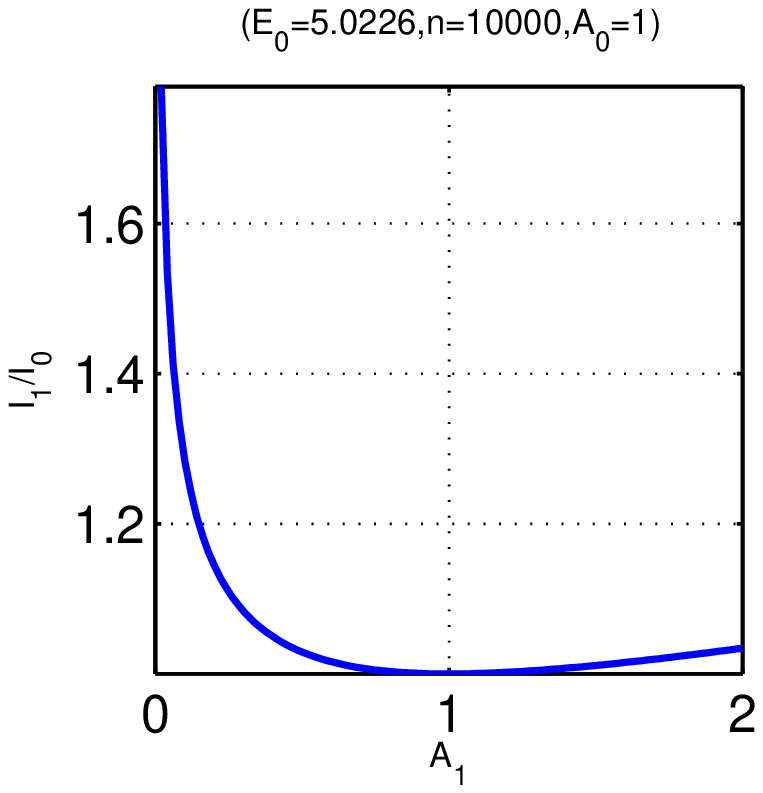}
\includegraphics[width=3.5cm]{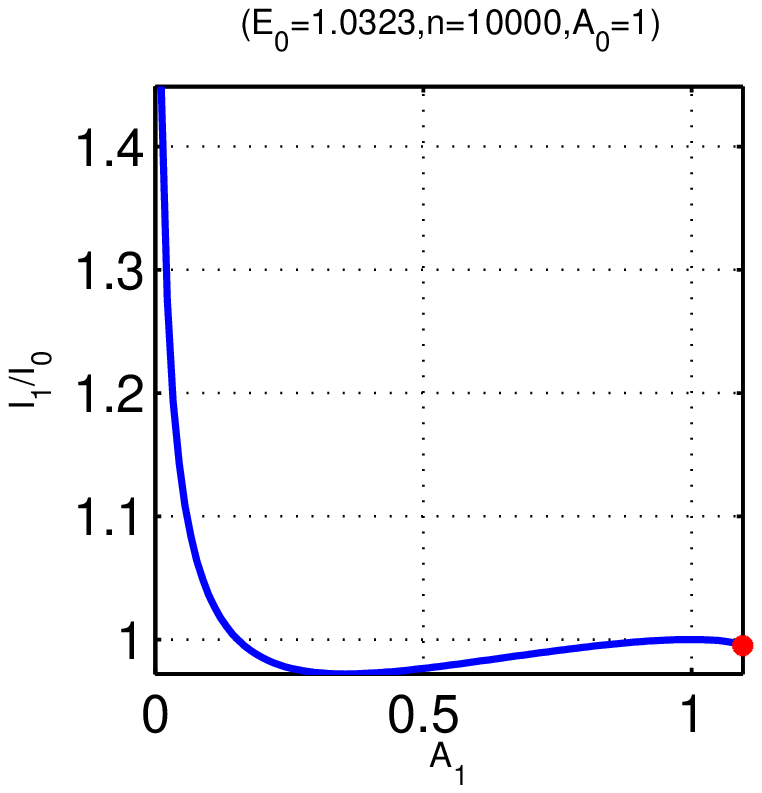}
\includegraphics[width=3.5cm]{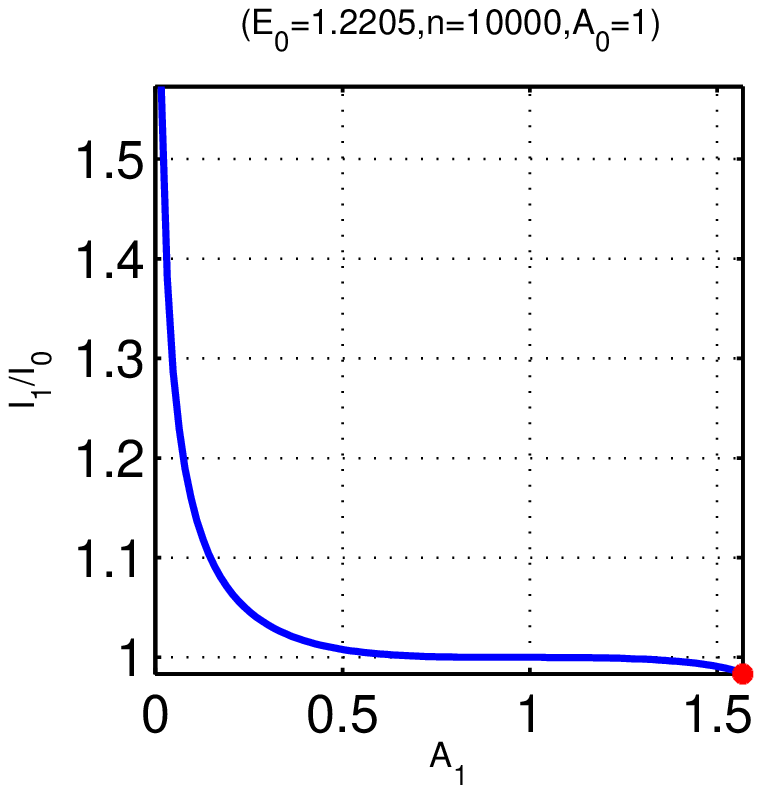}
\caption{Quartic double well potential: The action ratio
$I_{1}/I_{0}$ with $A_{0}=1$ and initial energy $E_0$ as a function
of $x=A_1$. $n$ is  the number of points in the initial ensemble 
and the bullet denotes the critical point at  which we cross the 
separatrix becoming trapped in one of
the two wells. Observe the behaviour at $A_1=1$, clockwise:
For (a) we find a minimum, in (b) we have a maximum,
and in (c) the inflection point.}
\label{IEquarpotF}
\end{figure}

The Hamiltonian is

\begin{equation}\label{qlhamiltonian}
H(q,p) = \frac{p^{2}}{2} - A \; f(q),\quad A>0,
\end{equation}
where the quartic double well potential is

\begin{equation}
V(q) = -Af(q) = A(q-1)^{2}(q+1)^{2}.
\end{equation}
We plot it in figure \ref{qdwpotfig}.
We start above the potential maximum $E_0 > A_0$, and make a 
kick $A_0\rightarrow A_1$.
The critical value $x_c$ of $x=A_1/A_0$ for the energy 
level $\langle E_1\rangle$  to stay outside the separatrix at $E=A_1$, 
namely $x\le x_c$,  is 

\begin{equation}
x_c = \frac{\pi I_0}{\pi I_0+ A_1T_0-E_0T_0}.
\end{equation}
We have calculated the action ratio $I_{1}/I_{0}$, assuming $A_0=1$,
as a function of $x=A_1$
using only numerical methods, and we find three different cases for the 
point $x=1$ in figure \ref{IEquarpotF}.
 For the energy $E=1.0323$ close to the potential
local maximum having level $V=1$ at $q=0$ we find a maximum of $F(x)$
at $x=1$, meaning the broken PR property, whilst at high energy
$E=5.0226$ we observe a local minimum at $x=1$, meaning that the
PR Property is satisfied. At intermediate energy $E=1.2205$ we
are close to an inflection point.

The examples of this section demonstrate that the PR property can
be broken if the potential has a single minimum but is not sufficiently 
smooth (if it is just $C^0$), whilst the class $C^1$ or higher is sufficient
for the PR property to hold. On the other hand, if the potential is
analytic but is not a single minimum potential, thus having a 
separatrix in the phase space, the PR property can be broken for energies
in the range near to the separatrix energy level. Moreover, the
potential can be analytic, with a single minimum, but has a region where
there is almost an inflection point, where the breaking
of PR property is observed for the energies in the range close to the
almost-inflection point.

\section{Discussion and conclusions} \label{Conclusion}

In this work we have analyzed the statistical properties of one degree
of freedom parametrically kicked Hamiltonian systems, which is the extreme 
case of fast time dependence, being the opposite extreme to an adiabatic, 
infinitely slow, changing. As such, the parametric kick behaviour is
a very good approximation for the behaviour of the systems under very
fast changes in the system parameters, within a time scale of less than one period
of oscillation, as has been demonstrated already by Papamikos
and Robnik \cite{PapRob2012}. The most natural ensemble, and the most important
one is the microcanonical ensemble, because if we have a large ensemble 
of identical systems with the same ("prepared") energy, and we do not have any
further information about them, the uniform distribution with respect
to the canonical angle ("the phases") is the most appropriate one. 
Our main interest is
in the value of the final average energy, just after the parametric kick,
and the value of the action, identical to the adiabatic invariant, or
identical to the area $\Omega$ inside the energy contour (divided by $2\pi$) 
in the phase space.
If the value of the adiabatic invariant at the average final
energy increases after the kick, we say that the system has the
PR property, following the conjecture put forward by Papamikos
and Robnik \cite{PapRob2012}. As discussed in section 2 this implies
increasing Gibbs entropy $S_G$ at the mean energy.

It turns out that the PR property is
satisfied in a vast variety of potentials in which we have proven
the validity of the conjecture by direct and rigorous calculations.
However, the PR Conjecture is not always satisfied. We have explored
exceptions and found that the PR property can be broken if the potential
is not sufficiently smooth (if it is $C^0$ only), or if it has
several local minima and maxima, implying existence of a separatrix,
or of several separatrices,  in
the phase space. Of course, these complications are not unexpected,
because the existence of a separatrix
in the phase space always
plays and important role, e.g. crossing a separatrix breaks the
validity of the adiabatic theorem \cite{Arnold}.
In energy ranges close to the separatrix, i.e.
stationary points of the potential, 
the PR property can be broken. Further study of the PR properties
in Hamiltonian systems with one and also with more degrees of freedom
seems to be very challenging and important. Our results suggest that
the following proposition might be true: A strictly convex $C^2$ potential
(thus with a single minimum) has the PR property. A proof of
this statement is lacking and left for the future.

We do believe that 
this research and the results are important in the statistical
mechanics of few body systems, and also for the large, macroscopical,
ensembles of identical {\em noninteracting} parametrically driven 
nonlinear  oscillators, as discussed in section 2. 
Further theoretical research is
in progress. Moreover, the research of the general statistical
behaviour of the energy distribution (and of other dynamical
variables), in the regimes between the ideal adiabatic variation
and the parametric kicking, is of great interest and importance.

\section*{Acknowledgements}

Financial support of the Slovenian Research Agency ARRS
under the grant P1-0306 is gratefully acknowledged. We also
thank the referees for constructive critical remarks and
very useful suggestions.

\section*{Appendix A:  Summary of the cases with or without  the PR property}

{\bf Examples of PR property valid for all energies}

\begin{enumerate}

\item Homogeneous power law kinetic energy and potential

\begin{equation}\label{homohamiltonian}
H(p,q)=\frac{p^{2n}}{2n} + \frac{Aq^{2m}}{2m},\quad A>0,\; m>0,\; n>0.
\end{equation}

\item Pendulum (for the average energy inside the separatrix - librations)

\begin{equation}\label{pendulumham}
H(q,p)= \frac{p^{2}}{2} - \Omega^{2} \cos q.
\end{equation}

\item Radial Kepler problem with zero angular momentum

\begin{equation}\label{keplerpotential}
H(r,p)= \frac{p^{2}}{2} - \frac{A}{r},\quad A>0.
\end{equation}

\item Radial Kepler problem with nonzero angular momentum

\begin{equation}\label{keplereffectivepotential}
H(r,p)= \frac{p^{2}}{2} - A \left(\frac{a}{r} -  \frac{M^{2}}{2r^{2}}\right),\quad A>0,\; a>0,\;M^{2}>0.
\end{equation}

\item Morse potential

\begin{equation}\label{morsepotential}
H(q,p)= \frac{p^{2}}{2} + A(e^{-2 \lambda r} - e^{- \lambda r}),\quad A>0,\quad \lambda>0.
\end{equation}

\item P\"oschl-Teller I potential

\begin{equation}  \label{pt1hamiltonian} 
H(q,p) = \frac{p^{2}}{2} + \frac{A}{\cos^{2} \left( \lambda q \right)},\quad \lambda>0,\; A>0.
\end{equation}

\item P\"oschl-Teller II potential

\begin{equation}\label{pt2hamiltonian}
H(q,p) = \frac{p^{2}}{2} - \frac{A}{\cosh^{2} \left(\lambda q\right)},\quad \lambda>0,\; A>0.
\end{equation}

\item Cosh potential

\begin{equation}\label{coshhamiltonian}
H(q,p) = \frac{p^{2}}{2} +  A \cosh \lambda q,\quad A>0.
\end{equation}

\item Quadratic-linear potential  $C^1$

\begin{equation}\label{qlhamiltonian}
H(q,p) = \frac{p^{2}}{2} - A \; f(q),\quad A>0,
\end{equation}
where

\begin{displaymath}
   f(q)=\left\{
     \begin{array}{lr}
       -\frac{\omega^{2} q^{2}}{2}, & \mbox{}  q\in[-q_{0},q_{0}] \\
- \om^2 q_0|q| + \frac{\om^2q_0^2}{2}, 
& \mbox{}  q \not\in [-q_{0},q_{0}].
     \end{array}
   \right.
\end{displaymath}

\item Quadratic-quartic potential $C^2$

\begin{equation}\label{qqhamiltonian}
H(q,p) = \frac{p^{2}}{2} - A \; f(q),\quad A>0,
\end{equation}
where
\begin{displaymath}
   f(q)=\left\{
     \begin{array}{lr}
       -\frac{\omega^{2} q^{2}}{2}, & \mbox{}  q\in[-q_{0},q_{0}] \\
       -\frac{\omega^{2}(q + 2 q_{0}\; {\rm sign}(q))^{4}}{108\;q_{0}^{2}} +
 \frac{\omega^{2} q_{0}^2}{4} , & \mbox{} 
q \not\in [-q_{0},q_{0}].
     \end{array}
   \right.
\end{displaymath}

\end{enumerate}

\noindent
{\bf Examples of violated PR property}

\begin{enumerate}

\item Harmonic oscillator in a box  ($C^0$, 
entirely convex, violation at $E\ge V(q_0)$ )

\begin{equation}\label{boxhamiltonian}
H(q,p) = \frac{p^{2}}{2} + V(q), 
\end{equation}
where

\begin{displaymath}
   V(q)=\left\{
     \begin{array}{lr}
       \frac{\omega^{2} q^{2}}{2}, & \mbox{}  q\in[-q_{0},q_{0}] \\
	   q = \infty , & \mbox{}  |q| \ge q_0
     \end{array}
   \right. .
\end{displaymath}

\item Sextic potential  (single minimum potential)

violation around $E=0$,  due
to the flatness near $|q|\approx 1$ (close to a stationary point)

\begin{equation} \label{qlhamiltonian}
H(q,p) = \frac{p^{2}}{2} - A \; f(q),\quad A>0,
\end{equation}
where the potential $V(q)=-Af(q)$ is given by

\begin{displaymath} 
f(q) = -(cq + (q-1)^{3})(cq + (q+1)^{3}), \; c\ge 0. 
\end{displaymath}

\item Quartic double well potential

PR property broken near the local maximum $E=A$.

\begin{equation}\label{qlhamiltonian}
H(q,p) = \frac{p^{2}}{2} - A \; f(q),\quad A>0,
\end{equation}
where the quartic double well potential is

\begin{displaymath}  
V(q) = -Af(q) = A(q-1)^{2}(q+1)^{2}.
\end{displaymath}

\end{enumerate}

\bibliography{myfer.bib}

\end{document}